\documentclass[lineno]{jfm}

\usepackage{graphicx}
\usepackage{newtxtext}
\usepackage{newtxmath}
\usepackage{natbib}
\usepackage{hyperref}
\usepackage{xcolor}
\usepackage{pifont}
\usepackage{tikz}
\usetikzlibrary{tikzmark}
\tikzset{mycircled/.style={circle,draw,inner sep=0.1em,line width=0.1em}}
\hypersetup{
    colorlinks = true,
    urlcolor   = blue,
    citecolor  = black,
}

\newcommand{\RomanNumeralCaps}[1]
\linenumbers

\title{Pressure-driven flow of the viscoelastic Oldroyd-B fluid in narrow non-uniform geometries: analytical results and comparison with simulations}

\author{Evgeniy Boyko\aff{1}
 \corresp{\email{eboyko@princeton.edu}}
 \and Howard A. Stone\aff{1}
 \corresp{\email{hastone@princeton.edu}}}

\affiliation{\aff{1}Department of Mechanical and Aerospace Engineering, Princeton University, Princeton, NJ 08544, USA}

\begin{document}
\maketitle

\begin{abstract}
We analyze the pressure-driven flow of a viscoelastic fluid in arbitrarily shaped, narrow channels and present a theoretical framework for calculating the relationship between the flow rate $q$ and pressure drop $\Delta p$. We utilize the Oldroyd-B model and first identify the characteristic scales and dimensionless parameters governing the flow in the lubrication limit. Employing a perturbation expansion in powers of the Deborah number ($De$), we provide analytical expressions for the velocity, stress, and the $q-\Delta p$ relation in the weakly viscoelastic limit up to $O(De^2)$. Furthermore, we exploit the reciprocal theorem derived by Boyko $\&$ Stone ({\it Phys. Rev. Fluids}, vol. 6, 2021, pp. L081301) to obtain the $q-\Delta p$ relation at the next order, $O(De^3)$, using only the velocity and stress fields at the previous orders. We validate our analytical results with two-dimensional numerical simulations of the Oldroyd-B fluid in a hyperbolic, symmetric contracting channel and find excellent agreement. For the flow-rate-controlled situation, both our theory and simulations reveal weak dependence of the velocity field on the Deborah number, so that the velocity can be approximated as Newtonian. In contrast to the velocity, the pressure drop strongly depends on the viscoelastic effects and decreases with $De$. Elucidating the relative importance of different terms in the momentum equation contributing to the pressure drop, we identify that a pressure drop reduction for narrow contracting geometries is primarily due to gradients in the viscoelastic shear stresses, while viscoelastic axial stresses have a minor effect on the pressure drop along the symmetry line.

\end{abstract}



\section{Introduction}

Pressure-driven flows of viscoelastic polymer solutions in narrow non-uniform geometries are widely encountered in industrial processes, such as molding and extrusion \citep{pearson,tadmor2013principles}, and in various applications ranging from microfluidic extensional rheometers \citep{ober2013microfluidic} to devices for subcutaneous drug administration, in which the liquid may exhibit non-Newtonian behavior \citep{allmendinger2014rheological,fischer2015calculation}. The complex rheological behavior of viscoelastic fluids affects the hydrodynamic features of such flows, including the relationship between the pressure drop $\Delta p$ across the channel and the flow rate $q$ even at low Reynolds number.

The dependence of the pressure drop on the flow rate of viscoelastic fluids at low Reynolds number has been studied extensively in various geometries. 
Table \ref{T1} shows a chronological selection of previous work on the $q-\Delta p$ relation for viscoelastic fluids in non-uniform geometries and clearly illustrates that the vast majority of the previous work involved numerical simulations and experimental measurements.

\begin{table}
\begin{center}
\def~{\hphantom{0}}
  \begin{tabular}{lccc}
 & Focus  & Geometry & Fluid/model  \\ \hline
 
 \citet{debbaut1988numerical} & Numer. & Planar and axisymmetric  & Oldroyd-B \\
&  & abrupt contraction & PTT/Giesekus  \\

\citet{keiller1993entry} & Numer. & Planar and axisymmetric  & Oldroyd-B \\
&  & abrupt contraction & FENE-CR  \\

\citet{szabo1997start} & Numer. & Axisymmetric abrupt  & FENE-CR  \\
&  & contraction-expansion &  \\

\citet{rothstein1999extensional}  & Exptl. & Axisymmetric abrupt  & Boger fluid   \\
\citet{rothstein2001axisymmetric} &   & contraction-expansion& \\

\citet{nigen2002viscoelastic}   & Exptl. & Three-dimensional and axisymmetric  & Boger fluid   \\
 &   &  abrupt contraction& \\

\citet{aboubacar2002highly} & Numer. & Planar abrupt contraction & Oldroyd-B/PTT \\

\citet{alves2003benchmark} & Numer. & Planar abrupt contraction & Oldroyd-B/PTT \\
 
 \citet{groisman2004microfluidic}   & Exptl. & Microfluidic rectifier    & Boger fluid   \\
 &   &  consisting of 43 tapered contractions & \\

\citet{binding2006contraction} & Numer. & Planar and axisymmetric abrupt & Oldroyd-B  \\
 &   & contraction, expansion& \\
  &   & and contraction-expansion& \\
 
 \citet{oliveira2007effect} & Numer. & Axisymmetric abrupt contraction & Oldroyd-B/PTT \\
  
\citet{aguayo2008excess} & Numer. & Planar and axisymmetric abrupt  & Oldroyd-B  \\
 &   & contraction and contraction-expansion& \\
  
  \citet{nguyen2008improvement}  & Exptl. & Microfluidic rectifier  consisting  & Boger fluid   \\
 &   &   of several tapered contractions & \\

  \citet{koppol2009anomalous} & Numer. & Axisymmetric abrupt & FENE-P \\
&  & contraction-expansion &  FENE bead- \\
&  & & spring chain \\

\citet{sousa2010efficient}  & Exptl. & Microfluidic rectifier  consisting of & Boger and  shear-  \\
 &   &   several tapered or hyperbolic  contractions & thinning fluids\\

  \citet{tamaddon2010predicting} & Numer. & Axisymmetric abrupt & FENE-CR \\
\citet{jahromi2011excess} &  & contraction-expansion& Oldroyd-B/PTT \\

  \citet{campo2011flow} & Exptl. & Three-dimensional hyperbolic  & Boger fluid \\
  &  & contraction followed by &  \\
&  &  an abrupt expansion &  \\

\citet{nystrom2012numerical} & Numer. & Axisymmetric abrupt & FENE-CR \\
 &   &  and hyperbolic contraction& \\

\citet{ober2013microfluidic} & Exptl. & Three-dimensional hyperbolic  & Boger and shear-\\
&  & contraction-expansion &  thinning fluids\\

  \citet{tamaddon2016predicting} & Numer. & Axisymmetric abrupt & FENE-CR \\
 &  & contraction-expansion& WM-FENE-CR \\

\citet{nystrom2016extracting} & Numer./ & Axisymmetric hyperbolic  & WM-FENE-CR \\
 & Exptl.  & contraction-expansion & \\
 
\citet{nystrom2017hyperbolic} & Numer./ & Axisymmetric hyperbolic contraction & Oldroyd-B  \\
 & Exptl.  &  & FENE-CR/PTT\\

 \citet{lopez2016numerical} & Numer. & Planar and axisymmetric  & WM-FENE-CR \\
 \citet{tamaddon2018modelling} &  & abrupt contraction &   \\
  
\citet{perez2019analytical} & Theor./ & Planar hyperbolic contraction & PTT with no  \\
 & Numer.  & & solvent contrib. \\
 
 Present work & Theor. & Planar  narrow slowly spatially & Oldroyd-B \\
 &   & varying geometries of arbitrary shape   & \\


 \end{tabular}
  \caption{Chronological selection of previous experimental, numerical, and theoretical papers on the flow rate$-$pressure drop relation for the low-Reynolds-number flows of viscoelastic fluids in non-uniform geometries.}
  \label{T1}
  \end{center}
\end{table}

The early studies on the flow rate$-$pressure drop relation have mainly investigated abrupt geometries such as contraction and contraction$-$expansion channels.
For such geometries, the two-dimensional (2-D) and axisymmetric numerical simulations with constitutive models, such as the Oldroyd-B model and finite-extensibility nonlinear elastic (FENE-CR) model introduced by \citet{chilcott1988creeping}, have generally predicted a reduction in the pressure drop with increasing Weissenberg
($Wi$) or Deborah ($De$) numbers, which are defined in $\mathsection$ \ref{Scaling}
\citep{keiller1993entry,szabo1997start,aboubacar2002highly,alves2003benchmark,binding2006contraction,oliveira2007effect,aguayo2008excess,tamaddon2010predicting,jahromi2011excess}.
The exceptions are simulations with small values of the finite extensibility parameter in the FENE-CR model that have reported an initial decrease in the pressure drop followed by a slight increase of the order of 10 $\%$ \citep{szabo1997start,tamaddon2010predicting,jahromi2011excess}. However, these predictions are in contrast with the experimental results of \citet{rothstein1999extensional,rothstein2001axisymmetric}, \citet{nigen2002viscoelastic}, and \citet{sousa2009three} for the flow of a polymer solution (Boger fluid) through abrupt axisymmetric contraction$-$expansion and contraction geometries that have reported a nonlinear increase in the pressure drop with the flow rate.  
Such an increase in the pressure drop was  observed also in  experimental studies on microfluidic rectifiers, which further showed the dependence of the flow rate$-$pressure drop relation on the flow direction \citep{groisman2004microfluidic,nguyen2008improvement,sousa2010efficient}. 

It is widely hypothesized that this discrepancy is attributed to the inability of the continuum level macroscale constitutive models, such as Oldroyd-B, FENE-CR, and the finite-extensibility nonlinear elastic model with the Peterlin approximation (FENE-P), to describe accurately the microscopic features of the polymer solutions \citep{owens2002computational,afonso2011dynamics}. As shown by \citet{koppol2009anomalous}, a mesoscopic level, micromechanical description, such as the bead-rod and bead-spring models, can be used to resolve, at least partially, this contradiction. For example, using the mesoscopic bead-spring chain model, \citet{koppol2009anomalous} showed an increase in the pressure drop for viscoelastic flow in an axisymmetric contraction$-$expansion geometry, which is in qualitative agreement with the experiments of \citet{rothstein1999extensional}. However,  \citet{koppol2009anomalous} were not able to observe such an agreement for simulations with the continuum FENE-P model, thus  indicating the advantage of mesoscopic over macroscopic simulations.

Nevertheless, it should be noted that, to date, the mesoscopic simulations are still computationally expensive and difficult to perform in complex geometries, requiring refined meshing and time-stepping for accurate viscoelastic predictions \citep{keunings2004micro,afonso2011dynamics,alves2021numerical}. Therefore, despite the limitations of a continuum approach, the vast majority of studies in non-Newtonian fluid mechanics still exploit the macroscopic constitutive equations, such as Oldroyd-B, FENE-CR, and FENE-P, which in principle can be modified to incorporate some microscopic features. For instance, Webster and co-workers proposed a new constitutive equation, which is the hybrid combination of White and Metzner \citep{white1963development} and FENE-CR models ~\citep[WM-FENE-CR, see, e.g.][]{jahromi2011excess,tamaddon2016predicting,webster2019enhanced}. Specifically, in this model, the deviatoric stress tensor  $\boldsymbol{\tau}$, which is the sum of the Newtonian  solvent and viscoelastic polymer  contributions, is obtained by multiplying the expression for  $\boldsymbol{\tau}$ from the FENE-CR model by a dissipative function $\phi(\dot{\varepsilon})$, usually taken as $\phi(\dot{\varepsilon})=1+(\lambda_{D}\dot{\varepsilon})^{2}$, where $\dot{\varepsilon}$ is the extension rate and $\lambda_{D}$ is an additional time constant. Such a model exhibits a constant shear viscosity, finite extensibility with a bounded extensional viscosity that reaches an ultimate plateau, and a first-normal stress-difference that has a weaker than quadratic dependence on the shear rate in the Oldroyd-B model. Using this hybrid WM-FENE-CR model, Webster and co-workers were able to achieve  quantitative agreement between their numerical predictions for the pressure drop and the earlier experiments of \citet{rothstein2001axisymmetric} and \citet{nigen2002viscoelastic}, where $\lambda_{D}$ and finite extensibility served as fitting parameters \citep{lopez2016numerical,tamaddon2016predicting,tamaddon2018modelling}.

After primarily focusing on abrupt contractions or contraction$-$expansions geometries a decade ago, hyperbolic symmetric channels with nearly constant extensional rates along the centerline have also received much attention, and several groups suggested to use of hyperbolic geometries for obtaining extensional properties of viscoelastic fluids through  $q-\Delta p$ measurements \citep{campo2011flow,ober2013microfluidic,keshavarz2016micro,nystrom2012numerical,nystrom2016extracting,nystrom2017hyperbolic,zografos2020viscoelastic}.  However, recently some researchers conjectured that, given the complex mixture of shear and extensional flow components in this geometry, it
is difficult to determine extensional viscosity directly from $q-\Delta p$ data \citep{james2016n1,hsiao2017passive}.

Recently, \citet{perez2019analytical} studied analytically and numerically the pressure-driven flow of a Phan-Thien$-$Tanner (PTT) fluid \citep{thien1977new,phan1978nonlinear} through a planar hyperbolic contraction. Using lubrication theory and neglecting the solvent contribution,  \citet{perez2019analytical} derived closed-form expressions for the non-dimensional velocity and pressure fields, which depend on the channel geometry and the product $\varepsilon_{PTT}Wi^2$, where $\varepsilon_{PTT}$ is the extensibility parameter of the PTT model.
Their results  predicted a decrease in the pressure drop with increasing $Wi$. However, such a reduction in the pressure drop arises due to shear-thinning effects of the PTT fluid, which are manifested when $Wi$ increases. Moreover, for $\varepsilon_{PTT}=0$, corresponding to the Oldroyd-B model, the solution of \citet{perez2019analytical} reduces to the Newtonian solution, which is independent of $Wi$. We have recently exploited the Lorentz reciprocal theorem and lubrication theory to derive a closed-form expression for the flow rate$-$pressure drop relation for complex fluids in narrow geometries, which holds for a wide class of non-Newtonian constitutive models \citep{boyko2021RT}. We showed the use of our theory to calculate analytically the first-order non-Newtonian correction for the $q-\Delta p$ relation for the viscoelastic second-order fluid and shear-thinning Carreau fluid, solely using the corresponding Newtonian solution and bypassing solution of the non-Newtonian flow problem. 

To the best of our knowledge, no analytical solution for the $q-\Delta p$ relation for constant shear-viscosity viscoelastic (Boger) fluids in narrow geometries has been reported in the literature, even for ``simple'' models such as Oldroyd-B and FENE-CR in the weakly viscoelastic limit. 
Such analytical solutions, however, are of fundamental importance as they may be used directly for comparison with experimental data and, in the case of discrepancy between the theory and experiments, may further provide insight into the cause of this disagreement and the adequacy of the constitutive model.

In this work, we provide a theoretical framework for calculating the flow rate$-$pressure drop relation of viscoelastic fluids in narrow channels of arbitrary shape. The present work presents analytical results for velocity and pressure fields and the $q-\Delta p$ relation for the Oldroyd-B model in the weakly viscoelastic limit. In subsequent work, we will analyze more complex constitutive models, incorporating additional microscopic features of polymer solutions. Our approach for obtaining analytical solutions for velocity and pressure is motivated by studies on thin films and lubrication problems \citep{tichy1996non,zhang2002surfactant,saprykin2007free,ahmed2021new}.  Such an approach relies on exploiting the narrowness of the geometry through the application of the lubrication approximation and a perturbation expansion in powers of the Deborah number $De$,  which is assumed to be small, $De\ll1$, and solving order by order, often resulting in cumbersome calculations at high orders. Instead, once the velocity at $O(De^2)$ is obtained, we use the reciprocal theorem, recently derived by \citet{boyko2021RT}, to calculate the pressure drop at $O(De^3)$, bypassing the detailed calculations of the viscoelastic flow problem at this order and relying only on the solution from previous orders. To validate the analytical results of our model, we perform two-dimensional finite-element numerical simulations with the Oldroyd-B model and find a good agreement between the theory and simulations, even for the cases when the hypotheses behind the lubrication approximation are not strictly satisfied. Given the recognized shortcomings of the Oldroyd-B and commonly used finite-extensibility nonlinear elastic (FENE) models in predicting the experimental observations for the flow rate$-$pressure drop relation, we are hopeful that the insights presented here may be useful in understanding possible physical or molecularly inspired modifications to the constitutive descriptions to improve future modeling and simulation efforts.

The paper is organised as follows. In $\mathsection$ \ref{PF}, we present the problem formulation and the dimensional governing equations and boundary conditions for the pressure-driven flow of the Oldroyd-B fluid. We further identify the characteristic scales and dimensionless parameters governing the flow and provide the non-dimensional governing equations. In $\mathsection$ \ref{LA}, we present a low-Deborah-number lubrication analysis and derive closed-form analytical solutions for the flow field and pressure drop up to $O(De^2)$. Exploiting the reciprocal theorem, in $\mathsection$ \ref{RT section} we calculate the pressure drop at $O(De^3)$, relying only on the solutions from previous orders. We present the results in $\mathsection$ \ref{Results}, including a comparison between the analytical predictions and the two-dimensional numerical simulations, finding excellent agreement between the two approaches. We conclude with a discussion of the results in $\mathsection$ \ref{CR}.

\section{Problem formulation and governing equations}\label{PF}

\begin{figure}
 \centerline{\includegraphics[scale=1]{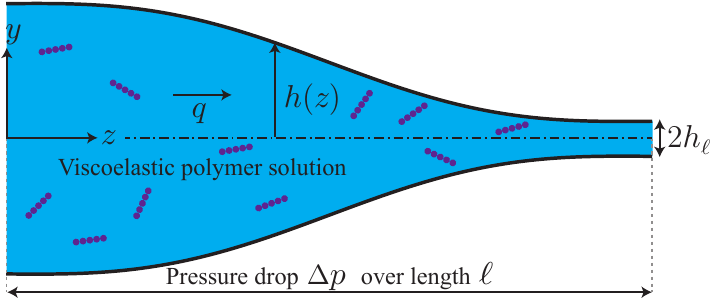}}
\caption{Schematic illustration of the two-dimensional configuration consisting of a spatially varying and
symmetric channel of height $2h(z)$ and length $\ell$. The channel contains a  viscoelastic dilute polymer solution steadily driven by an imposed
flow rate $q$, resulting in the pressure drop $\Delta p$.}\label{F1}
\end{figure}

We study the incompressible steady flow of a non-Newtonian
viscoelastic dilute polymer solution in a spatially varying
and symmetric two-dimensional channel of height $2h(z)$ and length $\ell$, where $h\ll\ell$. We assume that the imposed flow rate $q$ (per unit depth) induces the fluid motion with pressure distribution $p$ and velocity $\boldsymbol{u}=(u_{z},u_{y})$. Our primarily interest is to determine the resulting pressure drop $\Delta p$ over a streamwise distance $\ell$
for a given $q$. Figure \ref{F1} presents a schematic illustration of
the two-dimensional configuration and the coordinate system $(y,z)$,
whose $z$ axes lies in the symmetry midplane of the channel and $y$ 
is in the direction of the shortest dimension.


While throughout this work we consider steady and stable flows, it should be noted that the flow of viscoelastic fluids within non-uniform geometries may become unstable above a certain flow rate even at low Reynolds numbers due to the fluid's complex rheology~\citep{larson1992instabilities,shaqfeh1996purely,steinberg2021elastic,datta2021perspectives}. We consider low-Reynolds-number flows, so that the fluid inertia is negligible compared to viscous stresses. In this limit, the fluid motion is governed by the continuity and
momentum equations
\refstepcounter{equation}
$$
\boldsymbol{\nabla\cdot u}=0,\qquad\boldsymbol{\nabla\cdot\sigma}=\boldsymbol{0},\eqno{(\theequation{a,b})}\label{Continuity+Momentum}
$$where $\boldsymbol{\sigma}$ is the stress tensor given by
\begin{equation}
\boldsymbol{\sigma}=-p\boldsymbol{I}+2\eta_{s}\boldsymbol{E}+\boldsymbol{\tau}_{p}.\label{Stress tensor}
\end{equation}The first term on the right-hand side of (\ref{Stress tensor}) is
the pressure contribution, the second term is the viscous stress contribution
of Newtonian solvent with a constant viscosity $\eta_{s}$, where $\boldsymbol{E}=(\boldsymbol{\nabla}\boldsymbol{u}+(\boldsymbol{\nabla}\boldsymbol{u})^{\mathrm{T}})/2$ is the rate-of-strain tensor, and the last term, $\boldsymbol{\tau}_{p}$, is the polymer contribution to the stress
tensor.

In this work, we describe the viscoelastic behavior of the polymer solution using the Oldroyd-B constitutive model~\citep{bird1987dynamics1}. This is a widely used continuum model for Boger fluids, characterized by a constant shear viscosity. The Oldroyd-B equation can be derived from microscopic principles by modeling the polymer molecules as dumbbells, which follow a linear Hooke's law for the restoring force as they are  advected and stretched by the flow. In the Oldroyd-B model, the polymer contribution to the stress tensor
$\boldsymbol{\tau}_{p}$ can be expressed in the form~\citep{bird1987dynamics1,Intro_C_F,alves2021numerical}
\begin{equation}
\boldsymbol{\tau}_{p}=\frac{\eta_{p}}{\lambda}(\boldsymbol{A}-\boldsymbol{I}),\label{Polymer contribution to the stress tensor}
\end{equation}
where $\eta_{p}$ is the polymer contribution to the shear viscosity
at zero shear rate and $\lambda$ is the longest relaxation time of the
polymers. In (\ref{Polymer contribution to the stress tensor}), $\boldsymbol{A}$
is the conformation tensor of the dumbbells, which denotes the ensemble
average of the second moment of the dumbbell end-to-end vector $\boldsymbol{r}$
(scaled with its equilibrium value), $\boldsymbol{A}\equiv\left\langle \boldsymbol{rr}\right\rangle $,
and evolves at steady state according to~\citep{bird1987dynamics1,Intro_C_F,alves2021numerical}
\begin{equation}
\boldsymbol{u}\boldsymbol{\cdot}\boldsymbol{\nabla}\boldsymbol{A}-(\boldsymbol{\nabla}\boldsymbol{u})^{\mathrm{T}}\boldsymbol{\cdot}\boldsymbol{A}-\boldsymbol{A}\boldsymbol{\cdot}(\boldsymbol{\nabla}\boldsymbol{u})=-\frac{1}{\lambda}(\boldsymbol{A}-\boldsymbol{I}).\label{Evolution}
\end{equation}
Combining (\ref{Polymer contribution to the stress tensor}) and (\ref{Evolution}), we obtain an evolution equation for the polymer contribution to the stress tensor
$\boldsymbol{\tau}_{p}$, given at steady state as ~\citep{bird1987dynamics1,Intro_C_F,alves2021numerical},
\begin{equation}
\boldsymbol{\tau}_{p}+\lambda[\boldsymbol{u}\boldsymbol{\cdot}\boldsymbol{\nabla}\boldsymbol{\tau}_{p}-(\boldsymbol{\nabla}\boldsymbol{u})^{\mathrm{T}}\boldsymbol{\cdot}\boldsymbol{\tau}_{p}-\boldsymbol{\tau}_{p}\boldsymbol{\cdot}(\boldsymbol{\nabla}\boldsymbol{u})]=2\eta_{p}\boldsymbol{E}.
\label{Evolution tau_p}
\end{equation}
Using (\ref{Stress tensor}), (\ref{Polymer contribution to the stress tensor}), and (\ref{Evolution tau_p}), the stress tensor $\boldsymbol{\sigma}$ can be also expressed as 
\begin{equation}
\boldsymbol{\sigma}=-p\boldsymbol{I}+2\eta_{0}\boldsymbol{E}+\eta_{p}\boldsymbol{S},\label{Stress tensor 2}
\end{equation}
where $\eta_{0}=\eta_{s}+\eta_{p}$ is the total zero-shear-rate viscosity of the polymer solution and $\boldsymbol{S}$ is defined through  $\boldsymbol{\tau}_{p}=2\eta_{p}\boldsymbol{E}+\eta_{p}\boldsymbol{S}$, so that 
\begin{equation}
\boldsymbol{S}= -\frac{\lambda}{\eta_{p}}[\boldsymbol{u}\boldsymbol{\cdot}\boldsymbol{\nabla}\boldsymbol{\tau}_{p}-(\boldsymbol{\nabla}\boldsymbol{u})^{\mathrm{T}}\boldsymbol{\cdot}\boldsymbol{\tau}_{p}-\boldsymbol{\tau}_{p}\boldsymbol{\cdot}(\boldsymbol{\nabla}\boldsymbol{u})]\
 =-[\boldsymbol{u}\boldsymbol{\cdot}\boldsymbol{\nabla}\boldsymbol{A}-(\boldsymbol{\nabla}\boldsymbol{u})^{\mathrm{T}}\boldsymbol{\cdot}(\boldsymbol{A}-\boldsymbol{I})-(\boldsymbol{A}-\boldsymbol{I})\boldsymbol{\cdot}(\boldsymbol{\nabla}\boldsymbol{u})].\label{S tensor}
\end{equation}Substituting (\ref{Stress tensor 2}) into (\ref{Continuity+Momentum}) provides an alternative form of the governing equations  
\refstepcounter{equation}
$$
\boldsymbol{\nabla\cdot u}=0,\qquad\boldsymbol{\nabla}p=\eta_{0}\nabla^2\boldsymbol{u} + \eta_{p}\boldsymbol{\nabla\cdot S},\eqno{(\theequation{a,b})}\label{Continuity+Momentum with S}
$$which is convenient for assessing the viscoelastic effects on the the flow and pressure fields, where $\boldsymbol{S}$ is given in (\ref{S tensor}) and $\boldsymbol{A}$ evolves according to (\ref{Evolution}).

The governing equations (\ref{Continuity+Momentum})$-$(\ref{Continuity+Momentum with S})
are supplemented by the boundary conditions
\begin{equation}
u_{z}(h(z),z) =0, \quad u_{y}(h(z),z) =0, \quad \frac{\partial u_{z}}{\partial y}(0,z) = 0,\quad 2\int_{0}^{h(z)}u_{z}(y,z)\mathrm{d}y=q,\label{BC 2D}
\end{equation}corresponding, respectively, to no-slip and no-penetration 
along the walls of the channel, the symmetry boundary condition at the centerline, and the integral constraint stating that the total flow rate is prescribed.


\subsection{Scaling analysis and non-dimensionalization}\label{Scaling} 

In this work, we examine narrow configurations, in which $h(z)\ll\ell$, $h_{\ell}$ is the half-height at $x=\ell$, and $u_{c}=q/2h_{\ell}$ is the characteristic
velocity scale set by the cross-sectionally averaged velocity. Note that for the two-dimensional case, the flow rate $q$ is per unit depth. 

We introduce non-dimensional variables based on lubrication theory \citep{tichy1996non,zhang2002surfactant,saprykin2007free,ahmed2021new},
\begin{subequations}
\begin{equation}
Z=\frac{z}{\ell},\qquad Y=\frac{y}{h_{\ell}}, \qquad U_{z}=\frac{u_{z}}{u_{c}},\qquad U_{y}=\frac{u_{y}}{\epsilon u_{c}},\label{Non-dimensional variables 1}
\end{equation}
\begin{equation}
P=\frac{p}{\eta_{0}u_{c}\ell/h_{\ell}^{2}},\qquad\Delta P=\frac{\Delta p}{\eta_{0}u_{c}\ell/h_{\ell}^{2}},\quad\qquad H=\frac{h}{h_{\ell}}, \label{Non-dimensional variables 2}
\end{equation}
\begin{equation}
 \mathcal{T}_{p,zz}=\frac{\epsilon^{2}\ell}{\eta_{0}u_{c}}\tau_{p,zz},\qquad \tilde{A}_{zz}=\frac{\epsilon^{2}(A_{zz}-1)}{De},\qquad \mathcal{S}_{zz}=\frac{\epsilon^{2}\ell}{u_{c}De} S_{zz}, \label{Non-dimensional variables 3}
\end{equation}
\begin{equation}
 \mathcal{T}_{p,yz}=\frac{\epsilon\ell}{\eta_{0}u_{c}}\tau_{p,yz},\qquad \tilde{A}_{yz}=\frac{\epsilon A_{yz}}{De},\qquad \mathcal{S}_{yz}=\frac{\epsilon \ell}{u_{c} De}S_{yz}, \label{Non-dimensional variables 4}
\end{equation}
\begin{equation}
 \mathcal{T}_{p,yy}=\frac{\ell}{\eta_{0}u_{c}}\tau_{p,yy},\qquad \tilde{A}_{yy}=\frac{A_{yy}-1}{De}, \qquad  \mathcal{S}_{yy}=\frac{\ell}{u_{c} De}S_{yy},\label{Non-dimensional variables 5}
\end{equation}\label{ND variables}\end{subequations}where we have introduced the aspect ratio of the configuration, which is assumed to be small,
\begin{equation}
\epsilon=\frac{h_{\ell}}{\ell}\ll1,\label{epsilon}
\end{equation}
the viscosity ratios,
\begin{equation}
\tilde{\beta}=\frac{\eta_{p}}{\eta_{s}+\eta_{p}}=\frac{\eta_{p}}{\eta_{0}}\quad\mbox{and}\quad \beta=1-\tilde{\beta}=\frac{\eta_{s}}{\eta_{0}}\label{beta},
\end{equation}
and the Deborah and Weissenberg numbers, 
\begin{equation}
De=\frac{\lambda u_{c}}{\ell}\quad\mbox{and}\quad Wi=\frac{\lambda u_{c}}{h_{\ell}}.\label{Wi}
\end{equation}
The non-dimensional shape of the channel is denoted by $H(Z)$ and it will be an important parameter in our analysis.

Following \citet{ahmed2021new}, we define the Deborah number $De$ as the ratio of the polymer relaxation time, $\lambda$, to the residence time in the spatially non-uniform region, $\ell/u_{c}$, or alternatively, as the product of the relaxation time scale of the fluid and the characteristic extensional rate of the flow \citep[see also][]{tichy1996non,zhang2002surfactant,saprykin2007free}. The Weissenberg number $Wi$ is the product  of the relaxation time scale of the fluid and the characteristic shear rate of the flow, and is related to the Deborah number through $De=\epsilon Wi$ \citep{ahmed2021new}. We note that since we assume $\epsilon\ll1$, $De$ can be small while keeping $Wi =O(1)$.





\subsection{Governing equations in dimensionless form }

Using the non-dimensionalization (\ref{ND variables})$-$(\ref{Wi}), the governing equations (\ref{Continuity+Momentum})--(\ref{BC 2D}) take the form
\begin{subequations}
\begin{equation}
\frac{\partial U_{z}}{\partial Z}+\frac{\partial U_{y}}{\partial Y}=0,\label{Continuity ND}
\end{equation}
\begin{equation}
\frac{\partial P}{\partial Z}=\epsilon^{2}\frac{\partial^{2}U_{z}}{\partial Z^{2}}+\frac{\partial^{2}U_{z}}{\partial Y^{2}}+\tilde{\beta} De\frac{\partial \mathcal{S}_{zz}}{\partial Z}+\tilde{\beta} De\frac{\partial \mathcal{S}_{yz}}{\partial Y},\label{Momentum z ND}
\end{equation}
\begin{equation}
\frac{\partial P}{\partial Y}=\epsilon^{4}\frac{\partial^{2}U_{y}}{\partial Z^{2}}+\epsilon^{2}\frac{\partial^{2}U_{y}}{\partial Y^{2}}+\epsilon^{2}\tilde{\beta}De \frac{\partial  \mathcal{S}_{yz}}{\partial Z}+\epsilon^{2}\tilde{\beta} De\frac{\partial  \mathcal{S}_{yy}}{\partial Y},\label{Momentum y ND}
\end{equation}
\begin{equation}
De\left(U_{z}\frac{\partial A_{zz}}{\partial Z}+U_{y}\frac{\partial A_{zz}}{\partial Y}-2\frac{\partial U_{z}}{\partial Z}A_{zz}-2\frac{\partial U_{z}}{\partial Y}A_{yz}\right)-2\epsilon^{2}\frac{\partial U_{z}}{\partial Z}=-A_{zz},\label{Azz ND}
\end{equation}
\begin{equation}
De\left(U_{z}\frac{\partial A_{yy}}{\partial Z}+U_{y}\frac{\partial A_{yy}}{\partial Y}-2\frac{\partial U_{y}}{\partial Z}A_{yz}-2\frac{\partial U_{y}}{\partial Y}A_{yy}\right)-2\frac{\partial U_{y}}{\partial Y}=-A_{yy},\label{Ayy ND}
\end{equation}
\begin{equation}
De\left(U_{z}\frac{\partial A_{yz}}{\partial Z}+U_{y}\frac{\partial A_{yz}}{\partial Y}-\frac{\partial U_{y}}{\partial Z}A_{zz}-\frac{\partial U_{z}}{\partial Y}A_{yy}\right)-\epsilon^{2}\frac{\partial U_{y}}{\partial Z}-\frac{\partial U_{z}}{\partial Y}=-A_{yz},\label{Ayz ND}
\end{equation}
\label{ND}\end{subequations}subject to the boundary conditions
\begin{equation}
U_{z}(H(Z),Z)=0,\; U_{y}(H(Z),Z)=0,\; \frac{\partial U_{z}}{\partial Y}(0,Z) = 0,\; \int_{0}^{H(Z)}U_{z}(Y,Z)\mathrm{d}Y=1,\label{BC 2D ND}
\end{equation}
where we dropped tildes in the components of $\boldsymbol{A}$ for simplicity. From (\ref{Momentum y ND}), it follows that $P=P(Z)+O(\epsilon^{2})$,
i.e., the pressure is independent of $Y$ up to $O(\epsilon^{2})$,
consistent with the classical lubrication approximation. 

The explicit expressions for $\mathcal{S}_{zz}$ and $\mathcal{S}_{yz}$ appearing in (\ref{Momentum z ND}) are \begin{subequations}
\begin{equation}
\mathcal{S}_{zz}=-\left (U_{z}\frac{\partial A_{zz}}{\partial Z}+U_{y}\frac{\partial A_{zz}}{\partial Y}-2\dfrac{\partial U_{z}}{\partial Z}A_{zz}-2\dfrac{\partial U_{z}}{\partial Y}A_{yz}\right ),\label{Szz}
\end{equation}
\begin{equation}
\mathcal{S}_{yz}=-\left ( U_{z}\frac{\partial A_{yz}}{\partial Z}+U_{y}\frac{\partial A_{yz}}{\partial Y}-\dfrac{\partial U_{y}}{\partial Z}A_{zz}-\dfrac{\partial U_{z}}{\partial Y}A_{yy}\right ),\label{Syz}
\end{equation}\label{S}\end{subequations}and they are related to $A_{zz}$ and $A_{yz}$ through
\begin{subequations}
\begin{equation}
A_{zz}=De\mathcal{S}_{zz}+O(\epsilon^{2}),\qquad A_{yz}=De\mathcal{S}_{yz}+\dfrac{\partial U_{z}}{\partial Y}+O(\epsilon^{2}),\label{A-S relation}
\end{equation}
and to $\mathcal{T}_{p,zz}$ and $\mathcal{T}_{p,yz}$ through
\begin{equation}
\mathcal{T}_{p,zz}=\tilde{\beta}De\mathcal{S}_{zz}+O(\epsilon^{2}),\qquad \mathcal{T}_{p,yz}=\tilde{\beta}De\mathcal{S}_{yz}+\tilde{\beta}\dfrac{\partial U_{z}}{\partial Y}+O(\epsilon^{2}).\label{T-S relation}
\end{equation}
\end{subequations}

\section{Low-Deborah-number lubrication analysis}\label{LA}

In the previous section we obtained the non-dimensional equations (\ref{ND}), which are governed  by the three non-dimensional parameters: $\tilde{\beta}$, $De$, and $\epsilon^{2}$, where $\epsilon=h_{\ell}/\ell\ll1$.
In this section, we consider the weakly viscoelastic limit, $De\ll1$, and exploit the narrowness of the geometry, $\epsilon\ll1$, to derive analytical expressions for the velocity field and the $q-\Delta p$ relation for the pressure-driven flow of the Oldroyd-B model in a non-uniform channel of arbitrary shape $H(Z)$. We assume $\eta_{p}<\eta_{s}$, thus implying a dilute polymer solution with $\tilde{\beta}=\eta_{p}/\eta_{0}<1/2$ \citep{groisman1996couette,groisman2004microfluidic}. To this end, we seek solutions of the form
\begin{equation}
\left(\begin{array}{c}
U_{z}\\
U_{y}\\
P\\
A_{zz}\\
A_{yy}\\
A_{yz}
\end{array}\right)=\left(\begin{array}{c}
U_{z,0}\\
U_{y,0}\\
P_{0}\\
A_{zz,0}\\
A_{yy,0}\\
A_{yz,0}
\end{array}\right)+De\left(\begin{array}{c}
U_{z,1}\\
U_{y,1}\\
P_{1}\\
A_{zz,1}\\
A_{yy,1}\\
A_{yz,1}
\end{array}\right)+De^{2}\left(\begin{array}{c}
U_{z,2}\\
U_{y,2}\\
P_{2}\\
A_{zz,2}\\
A_{yy,2}\\
A_{yz,2}
\end{array}\right)+O(\epsilon^{2},De^{3}),\label{ND GE 2D}
\end{equation} and in the next subsections, we derive asymptotic expressions for the velocity field and the pressure drop up to $O(De^{2})$. In $\mathsection$ \ref{RT section}, we use reciprocal theorem to calculate the pressure drop at the next order, $O(De^{3})$.


  


\subsection{Leading-order solution}

Substituting (\ref{ND GE 2D}) into  (\ref{ND}) and considering the leading order in $De$, we obtain 
\begin{subequations}
\begin{equation}
\frac{\partial U_{z,0}}{\partial Z}+\frac{\partial U_{y,0}}{\partial Y}=0,\label{Continuity ND 2D LO}
\end{equation}
\begin{equation}
\frac{\mathrm{d}P_{0}}{\mathrm{d}Z}=\frac{\partial^{2}U_{z,0}}{\partial Y^{2}},\label{Momentum z ND 2D LO}
\end{equation}
\begin{equation}
\frac{\partial P_{0}}{\partial Y}=0,\label{Momentum y ND LO}
\end{equation}
\begin{equation}
A_{zz,0}=0,\label{Azz ND 2D LO}
\end{equation}
\begin{equation}
A_{yy,0}=2\dfrac{\partial U_{y,0}}{\partial Y},\label{Ayy ND 2D LO}
\end{equation}
\begin{equation}
A_{yz,0}=\dfrac{\partial U_{z,0}}{\partial Y},\label{Ayz ND 2D LO}
\end{equation}
\end{subequations}subject to the boundary conditions
\refstepcounter{equation}
$$
U_{z,0}(H(Z),Z)=0,\; U_{y,0}(H(Z),Z)=0,\; \frac{\partial U_{z,0}}{\partial Y}(0,Z) = 0,\; \int_{0}^{H(Z)}U_{z,0}(Y,Z)\mathrm{d}Y=1.\eqno{(\theequation{a-d})}\label{BC 2D ND LO}
$$As expected, at the leading order,  (\ref{Momentum z ND 2D LO}) reduces to the classical  momentum equation of the Newtonian fluid with a constant viscosity $\eta_{0}$. 

The solution of (\ref{Momentum z ND 2D LO}) using (\ref{BC 2D ND LO}$a$) and (\ref{BC 2D ND LO}$c$) is
\begin{equation}
U_{z,0}(Y,Z)=\frac{1}{2}\frac{\mathrm{d}P_{0}}{\mathrm{d}Z}\left(Y^{2}-H(Z)^{2}\right),\label{U(X,Y) LO}
\end{equation}
where the pressure gradient, which only depends on $Z$, follows from
applying the integral constraint (\ref{BC 2D ND LO}$d$),
\begin{equation}
\frac{\mathrm{d}P_{0}}{\mathrm{d}Z}=-\frac{3}{H(Z)^{3}}.\label{dP/dX LO e}
\end{equation}
The corresponding axial velocity distribution is then 
\begin{equation}
U_{z,0}(Y,Z)=\frac{3}{2}\frac{\left (H(Z)^{2}-Y^{2}\right )}{H(Z)^{3}}.\label{U(X,Y) LO e}
  \end{equation}
Substituting (\ref{U(X,Y) LO e}) into the continuity equation (\ref{Continuity ND 2D LO})
and using (\ref{BC 2D ND LO}$b$), yields
\begin{equation}
U_{y,0}(Y,Z)=\frac{3}{2}\frac{\mathrm{d}H(Z)}{\mathrm{d}Z}\frac{Y\left(H(Z)^{2}-Y^{2}\right)}{H(Z)^{4}},\label{V(X,Y) LO e}
\end{equation}
and thus the $yy$- and $yz$-components of
the conformation tensor at the leading-order depend on channel shape via
\begin{equation}
A_{yy,0}=\frac{3\mathrm{d}H(Z)}{\mathrm{d}Z}\frac{\left(-3Y^{2}+H(Z)^{2}\right)}{H(Z)^{4}},\qquad A_{yz,0}=-\frac{3Y}{H(Z)^{3}}.\label{Axx, Axy LO e}
\end{equation}
Finally, integrating (\ref{dP/dX LO e}) with respect to $Z$ from
0 to 1 provides the pressure drop at the leading order,
\begin{equation}
\Delta P_{0}=3\int_{0}^{1}\frac{\mathrm{d}Z}{H(Z)^{3}}.\label{Q-P relation LO}
\end{equation}

\subsection{First-order solution}\label{FO section}

At the first order, $O(De)$, the governing equations (\ref{ND})$-$(\ref{S})
yield\begin{subequations}
\begin{equation}
\frac{\partial U_{z,1}}{\partial Z}+\frac{\partial U_{y,1}}{\partial Y}=0,\label{Continuity ND 2D FO}
\end{equation}
\begin{equation}
\frac{\partial P_{1}}{\partial Z}=\frac{\partial^{2}U_{z,1}}{\partial Y^{2}}+\tilde{\beta}\left[\frac{\partial\mathcal{S}_{zz,0}}{\partial Z}+\frac{\partial\mathcal{S}_{yz,0}}{\partial Y}\right],\label{Momentum z ND 2D FO}
\end{equation}
\begin{equation}
\frac{\partial P_{1}}{\partial Y}=0,\label{Momentum y ND 2D FO}
\end{equation}
\begin{equation}
2\dfrac{\partial U_{z,0}}{\partial Y}A_{yz,0}=A_{zz,1},\label{Azz ND 2D FO}
\end{equation}
\begin{equation}
U_{z,0}\frac{\partial A_{yy,0}}{\partial Z}+U_{y,0}\frac{\partial A_{yy,0}}{\partial Y}-2\dfrac{\partial U_{y,0}}{\partial Z}A_{yz,0}-2\dfrac{\partial U_{y,0}}{\partial Y}A_{yy,0}-2\dfrac{\partial U_{y,1}}{\partial Y}=-A_{yy,1},\label{Ayy ND 2D FO}
\end{equation}
\begin{equation}
U_{z,0}\frac{\partial A_{yz,0}}{\partial Z}+U_{y,0}\frac{\partial A_{yz,0}}{\partial Y}-\dfrac{\partial U_{z,0}}{\partial Y}A_{yy,0}-\dfrac{\partial U_{z,1}}{\partial Y}=-A_{yz,1},\label{Ayz ND 2D FO}
\end{equation}
\begin{equation}
\mathcal{S}_{zz,0}=2\dfrac{\partial U_{z,0}}{\partial Y}A_{yz,0},\label{Szz0 2D}
\end{equation}
and
\begin{equation}
\mathcal{S}_{yz,0}=-U_{z,0}\frac{\partial A_{yz,0}}{\partial Z}-U_{y,0}\frac{\partial A_{yz,0}}{\partial Y}+\dfrac{\partial U_{0}}{\partial Y}A_{yy,0},\label{Syz0 2D}
\end{equation}
\end{subequations}where we have used (\ref{Azz ND 2D LO}) to simplify
(\ref{Azz ND 2D FO}), (\ref{Ayz ND 2D FO}), (\ref{Szz0 2D}), and (\ref{Syz0 2D}).

These governing equations are supplemented by the boundary conditions
\refstepcounter{equation}
$$
U_{z,1}(H(Z),Z)=0,\; U_{y,1}(H(Z),Z)=0,\; \frac{\partial U_{z,1}}{\partial Y}(0,Z) = 0,\; \int_{0}^{H(Z)}U_{z,1}(Y,Z)\mathrm{d}Y=0.\eqno{(\theequation{a-d})}\label{BC 2D ND FO}
$$
The last term on the right-hand side of (\ref{Momentum z ND 2D FO}) 
solely depends on the leading-order solution, and thus can be
explicitly calculated using (\ref{U(X,Y) LO e}), (\ref{V(X,Y) LO e}),
and (\ref{Axx, Axy LO e}) to yield,
\begin{equation}
\tilde{\beta}\left[\frac{\partial\mathcal{S}_{zz,0}}{\partial Z}+\frac{\partial\mathcal{S}_{yz,0}}{\partial Y}\right]=-\frac{18\tilde{\beta}}{H(Z)^{5}}\frac{\mathrm{d}H(Z)}{\mathrm{d}Z}=\frac{9\tilde{\beta}}{2}\frac{\mathrm{d}}{\mathrm{d}Z}\left ( \frac{1}{H(Z)^{4}}\right ).\label{f0 2D e}
\end{equation}
Next, integrating (\ref{Momentum z ND 2D FO}) twice with respect to $Y$, using 
(\ref{f0 2D e}), and applying the boundary conditions (\ref{BC 2D ND FO}$a$) and  (\ref{BC 2D ND FO}$c$), we obtain 
\begin{equation}
U_{z,1}(Y,Z)=\frac{1}{2}\frac{\mathrm{d}}{\mathrm{d}Z}\left (P_{1}-\frac{9}{2} \frac{\tilde{\beta}}{H(Z)^{4}}\right )\left(Y^{2}-H(Z)^{2}\right).\label{U FO e}
\end{equation}
To determine $\mathrm{d}P_{1}/\mathrm{d}Z$, we use the integral
constraint (\ref{BC 2D ND FO}$d$) to find
\begin{equation}
\frac{\mathrm{d}P_{1}}{\mathrm{d}Z}=\frac{9}{2}\tilde{\beta} \frac{\mathrm{d}}{\mathrm{d}Z}\left(\frac{1}{H(Z)^{4}}\right),\label{dP/dX FO e}
\end{equation}
and thus $U_{z,1}\equiv0$. From the continuity equation (\ref{Continuity ND 2D FO}),
it then follows that $U_{y,1}\equiv0$.

Integrating (\ref{dP/dX FO e}) with respect to
$Z$ from 0 to 1 provides the pressure drop at $O(De)$,
\begin{equation}
\Delta P_{1}=\frac{9}{2}\tilde{\beta}\left (\frac{1}{H(0)^{4}}-\frac{1}{H(1)^{4}}\right )=\frac{9}{2}\tilde{\beta}\left (\frac{H(1)^{4}-H(0)^{4}}{H(0)^{4}H(1)^{4}}\right ).\label{Q-P relation FO}
\end{equation}
We observe that, for the two-dimensional geometry, the integral
constraint (\ref{BC 2D ND FO}$d$) results in a balance between the gradients of the non-Newtonian stress contribution $\tilde{\beta}[(\partial\mathcal{S}_{zz}/\partial Z)+(\partial\mathcal{S}_{yz}/\partial Y)]$ and the pressure gradient $\mathrm{d}P_{1}/\mathrm{d}Z$ in the first-order momentum equation (\ref{Momentum z ND 2D FO}), such that there is no contribution to the velocity field at the first order. 



Since the velocity components
vanish at the first order, the components of the conformation tensor
at this order can be calculated using the leading-order velocity field,\begin{subequations}
\begin{equation}
A_{zz,1}=\frac{18Y^{2}}{H(Z)^{6}},\label{Axx1}
\end{equation}
\begin{equation}
A_{yz,1}=18\frac{\mathrm{d}H(Z)}{\mathrm{d}Z}\frac{Y\left(H(Z)^{2}-2Y^{2}\right)}{H(Z)^{7}},\label{Axy1}
\end{equation}
\begin{equation}
A_{yy,1}=\frac{9}{2}\frac{4\left(-2Y^{2}+H(Z)^{2}\right)^{2}H'(Z)^{2}-H(Z)H''(Z)\left(Y^{2}-H(Z)^{2}\right)^{2}}{H(Z)^{8}},\label{Ayy1}
\end{equation}
\end{subequations}where primes indicate derivatives with respect to $Z$. 


\subsection{Second-order solution}

At the second order, $O(De^{2})$, the governing equations (\ref{ND})$-$(\ref{S}) take the form\begin{subequations}
\begin{equation}
\frac{\partial U_{z,2}}{\partial Z}+\frac{\partial U_{y,2}}{\partial Y}=0,\label{Continuity ND 2D SO}
\end{equation}
\begin{equation}
\frac{\mathrm{d}P_{2}}{\mathrm{d}Z}=\frac{\partial^{2}U_{z,2}}{\partial Y^{2}}+\tilde{\beta}\left[\frac{\partial\mathcal{S}_{zz,1}}{\partial Z}+\frac{\partial\mathcal{S}_{yz,1}}{\partial Y}\right],\label{Momentum z ND 2D SO}
\end{equation}
\begin{equation}
\frac{\partial P_{2}}{\partial Y}=0,\label{Momentum y ND 2D SO}
\end{equation}
\begin{equation}
U_{z,0}\frac{\partial A_{zz,1}}{\partial Z}+U_{y,0}\frac{\partial A_{zz,1}}{\partial Y}-2\dfrac{\partial U_{z,0}}{\partial Z}A_{zz,1}-2\dfrac{\partial U_{z,0}}{\partial Y}A_{yz,1}=-A_{zz,2},\label{Azz ND 2D SO}
\end{equation}
\begin{equation}
U_{z,0}\frac{\partial A_{yy,1}}{\partial Z}+U_{y,0}\frac{\partial A_{yy,1}}{\partial Y}-2\dfrac{\partial U_{y,0}}{\partial Z}A_{yz,1}-2\dfrac{\partial U_{y,0}}{\partial Y}A_{yy,1}-2\dfrac{\partial U_{y,2}}{\partial Y}=-A_{yy,2},\label{Ayy ND 2D SO}
\end{equation}
\begin{equation}
U_{z,0}\frac{\partial A_{yz,1}}{\partial Z}+U_{y,0}\frac{\partial A_{yz,1}}{\partial Y}-\dfrac{\partial U_{y,0}}{\partial Z}A_{zz,1}-\dfrac{\partial U_{z,0}}{\partial Y}A_{yy,1}-\dfrac{\partial U_{z,2}}{\partial Y}=-A_{yz,2},\label{Ayz ND 2D SO}
\end{equation}
\begin{equation}
\mathcal{S}_{zz,1}=-U_{z,0}\frac{\partial A_{zz,1}}{\partial Z}-U_{y,0}\frac{\partial A_{zz,1}}{\partial Y}+2\dfrac{\partial U_{z,0}}{\partial Z}A_{zz,1}+2\dfrac{\partial U_{z,0}}{\partial Y}A_{yz,1},\label{Szz1 2D}
\end{equation}
\begin{equation}
\mathcal{S}_{yz,1}=-U_{z,0}\frac{\partial A_{yz,1}}{\partial Z}-U_{y,0}\frac{\partial A_{yz,1}}{\partial Y}+\dfrac{\partial U_{y,0}}{\partial Z}A_{zz,1}+\dfrac{\partial U_{z,0}}{\partial Y}A_{yy,1},\label{Syz1 2D}\end{equation}\label{SO 2D}\end{subequations}where we have used the fact that $U_{z,1}\equiv0$ and $U_{y,1}\equiv0$. The governing equations (\ref{SO 2D}) are supplemented by the boundary conditions
\refstepcounter{equation}
$$
U_{z,2}(H(Z),Z)=0,\; U_{y,2}(H(Z),Z)=0,\; \frac{\partial U_{z,2}}{\partial Y}(0,Z) = 0,\; \int_{0}^{H(Z)}U_{z,2}(Y,Z)\mathrm{d}Y=0.\eqno{(\theequation{a-d})}\label{BC 2D ND SO}
$$
The last term on the right-hand side of (\ref{Momentum z ND 2D SO}) 
solely depends on the leading- and first-order solutions, and
thus can be calculated using (\ref{Szz1 2D}$)-($\ref{Syz1 2D}),
\begin{equation}
\tilde{\beta}\left[\frac{\partial\mathcal{S}_{zz,1}}{\partial Z}+\frac{\partial\mathcal{S}_{yz,1}}{\partial Y}\right]=\frac{81\tilde{\beta}}{2}\frac{\left(Y^{2}-H(Z)^{2}\right)\left[4H(Z)H'(Z)^{2}+H''(Z)\left(Y^{2}-H(Z)^{2}\right)\right]}{H(Z)^{10}}.\label{f1 2D e}
\end{equation}
Integrating (\ref{Momentum z ND 2D SO}) twice with respect to $Y$, using 
(\ref{f1 2D e}),
and applying the boundary conditions (\ref{BC 2D ND SO}$a$) and (\ref{BC 2D ND SO}$c$),
we obtain 
\begin{equation}
U_{z,2}=(Y^{2}-H^{2})\left (\frac{1}{2}\frac{\mathrm{d}P_{2}}{\mathrm{d}Z}-\frac{27\tilde{\beta}}{20}\frac{10HH'^{2}\left(Y^{2}-5H^{2}\right)+H''\left(Y^{4}-4(YH)^{2}+11H^{4}\right)}{H^{10}}\right).\label{U2}
\end{equation}
To determine $\mathrm{d}P_{2}/\mathrm{d}Z$, we use the integral constraint
(\ref{BC 2D ND SO}$c$), to find
\begin{equation}
\frac{\mathrm{d}P_{2}}{\mathrm{d}Z}=-\frac{324}{35}\tilde{\beta}\left (\frac{14H'(Z)^{2}}{H(Z)^{7}}-\frac{3H''(Z)}{H(Z)^{6}}\right ).\label{dP/dX SO e}
\end{equation}
Integrating (\ref{dP/dX SO e}) with respect to $Z$ from 0 to 1 yields the pressure drop at $O(De^{2}),$
\begin{equation}
\Delta P_{2}=\frac{324}{35}\tilde{\beta}\int_{0}^{1}\left (\frac{14H'(Z)^{2}}{H(Z)^{7}}-\frac{3H''(Z)}{H(Z)^{6}}\right)\mathrm{d}Z.\label{Q-P relation SO}
\end{equation}
For a given flow rate $q$, we have determined the dimensionless pressure
drop $\Delta P=\Delta p/(\eta_{0}q\ell/2h_{\ell}^{3})$ as a function
of the shape function $H(Z)$, the viscosity ratio $\tilde{\beta}$ and the
Deborah number $De$,
\begin{equation}
\Delta P=\Delta P_{0}(H(Z))+De\Delta P_{1}(\tilde{\beta},H(Z))+De^{2}\Delta P_{2}(\tilde{\beta},H(Z))+O(\epsilon^{2},De^{3}),\label{Pressure drop}
\end{equation}
where the expressions for $\Delta P_{0}$, $\Delta P_{1}$, and $\Delta P_{2}$
are given in (\ref{Q-P relation LO}), (\ref{Q-P relation FO}), and
(\ref{Q-P relation SO}), respectively. 

We note that once $U_{z,2}(Y,Z)$ is determined from (\ref{U2}) and
(\ref{dP/dX SO e}), $U_{y,2}(Y,Z)$ can be found using the continuity
(\ref{Continuity ND 2D SO}) and (\ref{BC 2D ND SO}$b$). Furthermore, the components of the conformation
tensor at this order can be calculated from (\ref{Azz ND 2D SO})$-$(\ref{Ayz ND 2D SO}).
While the resulting expressions are readily found  using \textsc{Mathematica}, they are rather lengthy and thus not presented
here.

\section{Reciprocal theorem for the flow of an Oldroyd-B fluid in narrow geometries}\label{RT section}

In this section, we exploit the reciprocal theorem for complex
fluids in narrow geometries, recently derived by \citet{boyko2021RT}, to calculate the flow rate$-$pressure drop relation, bypassing the detailed calculations of the viscoelastic  flow problem. In particular, we show that the reciprocal theorem allows one to obtain the $q-\Delta p$ relation at the next order, $O(De^3)$, relying only on the solutions from previous orders. For completeness, we present the governing
equations and the key relations derived in \citet{boyko2021RT}, adapted to the notation in this paper.

Let $\hat{\boldsymbol{u}}$ and $\hat{\boldsymbol{\sigma}}$ denote,
respectively, the velocity and stress fields corresponding to the
solution of the Stokes equations in the same domain with the constant viscosity $\eta_{0}$. The corresponding governing equations
are 
\refstepcounter{equation}
$$
\boldsymbol{\nabla\cdot}\hat{\boldsymbol{u}}=0,\qquad\boldsymbol{\nabla\cdot}\hat{\boldsymbol{\sigma}}=\boldsymbol{0}\quad\mathrm{with\quad}\hat{\boldsymbol{\sigma}}=-\hat{p}\boldsymbol{I}+2\eta_{0}\hat{\boldsymbol{E}}.\eqno{(\theequation{a,b})}\label{Continuity+Momentum solvent}
$$The reciprocal theorem states that two flows $(\boldsymbol{u},\boldsymbol{\sigma})$ and $(\hat{\boldsymbol{u}},\hat{\boldsymbol{\sigma}})$, governed by (\ref{Continuity+Momentum with S}) and (\ref{Continuity+Momentum solvent}), satisfy \citep[see][]{boyko2021RT},
\begin{equation}
\int_{S_{0}}\boldsymbol{n}\boldsymbol{\cdot}\boldsymbol{\sigma}\boldsymbol{\cdot}\hat{\boldsymbol{u}}\mathrm{d}S+\int_{S_{\ell}}\boldsymbol{n}\boldsymbol{\cdot}\boldsymbol{\sigma}\boldsymbol{\cdot}\hat{\boldsymbol{u}}\mathrm{d}S-\int_{S_{0}}\boldsymbol{n}\boldsymbol{\cdot}\hat{\boldsymbol{\sigma}}\boldsymbol{\cdot}\boldsymbol{u}\mathrm{d}S-\int_{S_{\ell}}\boldsymbol{n}\boldsymbol{\cdot}\hat{\boldsymbol{\sigma}}\boldsymbol{\cdot}\boldsymbol{u}\mathrm{d}S=\eta_{p}\int_{\mathcal{V}}\boldsymbol{S}\boldsymbol{:}\hat{\boldsymbol{E}}\mathrm{d}\mathcal{V},\label{Reciprocal theorem}
\end{equation}
where $\mathcal{V}$ is the entire fluid volume bounded by the surface of the top
and bottom walls $S_{w}$, and the surfaces at the inlet and outlet
$S_{0}$ and $S_{\ell}$ at $z=0$ and $z=\ell$, respectively, and $\boldsymbol{n}$ is the unit outward normal to $S_{0,\,\ell}$. Note that the integrals over the walls $S_{w}$
vanish since there $\boldsymbol{u}=\hat{\boldsymbol{u}}=\boldsymbol{0}$.


Using the scaling analysis and (\ref{Stress tensor 2}), (\ref{ND variables}), and (\ref{Continuity+Momentum solvent}), the terms $\eta_{p}\boldsymbol{S}\boldsymbol{:}\hat{\boldsymbol{E}}$,
$\boldsymbol{n}\boldsymbol{\cdot}\boldsymbol{\sigma}\boldsymbol{\cdot}\hat{\boldsymbol{u}}$,
and $\boldsymbol{n}\boldsymbol{\cdot}\boldsymbol{\sigma}\boldsymbol{\cdot}\hat{\boldsymbol{u}}$,
appearing in (\ref{Reciprocal theorem}), are approximately:\begin{subequations}
\begin{equation}
 \begin{array}{ccc}
\displaystyle \eta_{p}\boldsymbol{S}\boldsymbol{:}\hat{\boldsymbol{E}}=\frac{\eta_{p}Deu_{c}^{2}}{h_{\ell}^{2}}\left (\mathcal{S}_{zz}\frac{\partial\hat{U}_{z}}{\partial Z}+\mathcal{S}_{yz}\frac{\partial\hat{U}_{z}}{\partial Y}+O(\epsilon^{2})\right ),
 \end{array}\label{RT Scaling 1}\end{equation}
\begin{equation}
\boldsymbol{n}\boldsymbol{\cdot}\boldsymbol{\sigma}\boldsymbol{\cdot}\hat{\boldsymbol{u}}=\mp\frac{\eta_{0}u_{c}^{2}\ell}{h_{\ell}^{2}}\left[\left(-P+\tilde{\beta}De\mathcal{S}_{zz}\right)\hat{U}_{z}+O(\epsilon^{2})\right]_{Z=0,\,1},\label{RT Scaling 2}
\end{equation}
\begin{equation}
\boldsymbol{n}\boldsymbol{\cdot}\hat{\boldsymbol{\sigma}}\boldsymbol{\cdot}\boldsymbol{u}=\mp\frac{\eta_{0}u_{c}^{2}\ell}{h_{\ell}^{2}}\left[-\hat{P}U_{z}+O(\epsilon^{2})\right]_{Z=0,\,1},\label{RT Scaling 3}
\end{equation}\label{RT Scaling}\end{subequations}
where the minus sign in (\ref{RT Scaling 2}) and (\ref{RT Scaling 3})
corresponds to $S_{0}$ and the plus sign corresponds to $S_{\ell}$.
Substituting (\ref{RT Scaling}) into (\ref{Reciprocal theorem}),
we obtain 
\begin{eqnarray}
&   & \int_{0}^{H(0)}\left[(P-\tilde{\beta} De\mathcal{S}_{zz})\hat{U}_{z}-\hat{P}U_{z}\right]_{Z=0}\mathrm{d}Y-\int_{0}^{H(1)}\left[(P-\tilde{\beta} De\mathcal{S}_{zz})\hat{U}_{z}-\hat{P}U_{z}\right]_{Z=1}\mathrm{d}Y \nonumber\\
  &  & \mbox{\hspace{15mm}} =\tilde{\beta} De\int_{0}^{1}\int_{0}^{H(Z)}\left(\mathcal{S}_{zz}\frac{\partial\hat{U}_{z}}{\partial Z}+\mathcal{S}_{yz}\frac{\partial\hat{U}_{z}}{\partial Y}\right)\mathrm{d}Y\mathrm{d}Z+O(\epsilon^{2}),\label{RT 4 2D}
\end{eqnarray}
where $H(Z)$ is the non-dimensional shape of the channel. Noting that $P=P(Z)+O(\epsilon^{2})$, $\hat{P}=\hat{P}(Z)+O(\epsilon^{2})$,
and $\int_{0}^{H(Z)}U_{z}\mathrm{d}Y=\int_{0}^{H(Z)}\hat{U}_{z}\mathrm{d}Y=1$,
and defining $\Delta  P=P(0)-P(1)$ and $\Delta \hat{P}=\hat{P}(0)-\hat{P}(1)$,
(\ref{RT 4 2D}) simplifies to 
\begin{eqnarray}
\Delta  P &=& \Delta \hat{P}+\tilde{\beta} De\int_{0}^{H(0)}\left[\mathcal{S}_{zz}\hat{U}_{z}\right]_{Z=0}\mathrm{d}Y-\tilde{\beta} De\int_{0}^{H(1)}\left[\mathcal{S}_{zz}\hat{U}_{z}\right]_{Z=1}\mathrm{d}Y\nonumber\\
  & + & \tilde{\beta} De\int_{0}^{1}\int_{0}^{H(Z)}\left(\mathcal{S}_{zz}\frac{\partial\hat{U}_{z}}{\partial Z}+\mathcal{S}_{yz}\frac{\partial\hat{U}_{z}}{\partial Y}\right)\mathrm{d}Y\mathrm{d}Z+O(\epsilon^{2}),\label{RT 5 2D}
\end{eqnarray}
where the solution of the corresponding Newtonian problem is obtained from (\ref{U(X,Y) LO e}), (\ref{V(X,Y) LO e}), and (\ref{Q-P relation LO}) as
\begin{equation}
\Delta\hat{P}=3\int_{0}^{1}\frac{\mathrm{d}Z}{H(Z)^{3}},\quad\hat{U}_{z}=\frac{3}{2}\frac{H(Z)^{2}-Y^{2}}{H(Z)^{3}},\quad \hat{U}_{y}=\frac{3}{2}\frac{\mathrm{d}H(Z)}{\mathrm{d}Z}\frac{Y\left(H(Z)^{2}-Y^{2}\right)}{H(Z)^{4}}.\label{Solution-Newtonian problem 2D}
\end{equation}
Equation (\ref{RT 5 2D}) indicates that the pressure drop of the viscoelastic flow of an Oldroyd-B fluid in a narrow channel consists of four contributions. The first term on the right-hand
side of (\ref{RT 5 2D}) represents the Newtonian contribution to the
pressure drop. The second and third terms represent the contribution of the viscoelastic normal stress of the complex fluid at the inlet and outlet of the channel. Finally, the last term represents the viscoelastic contribution due to elongational $(\mathcal{S}_{zz}\partial\hat{U}_{z}/\partial Z)$
and shearing $(\mathcal{S}_{yz}\partial\hat{U}_{z}/\partial Y)$ effects within the fluid domain ${\mathcal{V}}$.

Furthermore, (\ref{RT 5 2D}) clearly shows that the pressure drop depends on the $\mathcal{S}_{zz}$ and $\mathcal{S}_{yz}$, and thus, generally,
requires the solution of the nonlinear viscoelastic problem. However,
in the weakly viscoelastic limit, corresponding to $De\ll1$, the reciprocal theorem
(\ref{RT 5 2D}) allows one to determine the pressure drop at the
current order only with the knowledge of the solution of the Newtonian
problem and previous orders. For example, we can
determine $\Delta P_{1}$ with the knowledge of the solution of the
Newtonian problem and the leading-order solution. Similarly, we can determine $\Delta P_{2}$ with the knowledge of the solution of the Newtonian problem and the leading- and first-order solutions of the viscoelastic problem. We note that our analysis assumes only negligible fluid inertia, a shallow geometry, $\epsilon\ll1$, and the weakly viscoelastic limit, $De\ll1$, while allowing $Wi$ to be $O(1)$.

In the next sections, we illustrate the use of the reciprocal theorem (\ref{RT 5 2D}) and provide closed-form analytical expressions for the pressure drop of an Oldroyd-B fluid up to $O(De^3)$ for two-dimensional geometries.


\subsection{Expression for the dimensionless  pressure drop at the first order}

Substituting (\ref{ND GE 2D}) into  (\ref{RT 5 2D}) and considering the first order, $O(De)$, we obtain 
\begin{eqnarray}
\Delta P_{1} &= &\tilde{\beta}\int_{0}^{H(0)}\left[\mathcal{S}_{zz,0}\hat{U}_{z}\right]_{Z=0}\mathrm{d}Y-\tilde{\beta}\int_{0}^{H(1)}\left[\mathcal{S}_{zz,0}\hat{U}_{z}\right]_{Z=1}\mathrm{d}Y \nonumber\\
  & & +\tilde{\beta}\int_{0}^{1}\int_{0}^{H(Z)}\left(\mathcal{S}_{zz,0}\frac{\partial\hat{U}_{z}}{\partial Z}+\mathcal{S}_{yz,0}\frac{\partial\hat{U}_{z}}{\partial Y}\right)\mathrm{d}Y\mathrm{d}Z.\label{RT FO 1 2D}
\end{eqnarray}Equation (\ref{RT FO 1 2D}) indicates that the first-order pressure
drop $\Delta  P_{1}$ can be calculated with the knowledge of the solutions
of the Newtonian problem and the leading-order viscoelastic problem.
Using the expressions for $\mathcal{S}_{zz,0}$ and $\mathcal{S}_{yz,0}$,
given in (\ref{Szz0 2D}) and (\ref{Syz0 2D}), we obtain
\begin{equation}
\Delta P_{1}=\frac{18}{5}\tilde{\beta}\left (\frac{1}{H(0)^{4}}-\frac{1}{H(1)^{4}}\right )+\frac{9}{10}\tilde{\beta}\left (\frac{1}{H(0)^{4}}-\frac{1}{H(1)^{4}}\right )=\frac{9}{2}\tilde{\beta}\left(\frac{1}{H(0)^{4}}-\frac{1}{H(1)^{4}}\right ),\label{RT FO 3 2D}
\end{equation}
which is exactly (\ref{Q-P relation FO}).

\subsection{Expression for the dimensionless pressure drop at the second order}

At the second order, $O(De^{2})$, we have
\begin{eqnarray}
\Delta P_{2} &= &\tilde{\beta}\int_{0}^{H(0)}\left[\mathcal{S}_{zz,1}\hat{U}_{z}\right]_{Z=0}\mathrm{d}Y-\tilde{\beta}\int_{0}^{H(1)}\left[\mathcal{S}_{zz,1}\hat{U}_{z}\right]_{Z=1}\mathrm{d}Y \nonumber\\
  & & +\tilde{\beta}\int_{0}^{1}\int_{0}^{H(Z)}\left(\mathcal{S}_{zz,1}\frac{\partial\hat{U}_{z}}{\partial Z}+\mathcal{S}_{yz,1}\frac{\partial\hat{U}_{z}}{\partial Y}\right)\mathrm{d}Y\mathrm{d}Z.\label{RT SO 1 2D}
\end{eqnarray}
The second-order pressure
drop $\Delta P_{2}$ solely depends on the solution of the Newtonian
problem and the leading- and first-order viscoelastic problems. Using the expressions for $\mathcal{S}_{zz,1}$ and $\mathcal{S}_{yz,1}$, given in (\ref{Szz1 2D}) and (\ref{Syz1 2D}), and (\ref{Solution-Newtonian problem 2D}), we obtain 
\begin{equation}
\Delta P_{2}=\frac{648}{35}\tilde{\beta}\left(\frac{H'(0)}{H(0)^{6}}-\frac{H'(1)}{H(1)^{6}}\right)+\frac{324}{35}\tilde{\beta}\int_{0}^{1}\left (\frac{2H'(Z)^{2}}{H(Z)^{7}}-\frac{H''(Z)}{H(Z)^{6}}\right )\mathrm{d}Z,\label{RT SO 2 2D}
\end{equation}
which can be rewritten as,
\begin{equation}
\Delta P_{2}=\frac{324}{35}\tilde{\beta}\int_{0}^{1}\left[\frac{14H'(Z)^{2}}{H(Z)^{7}}-\frac{3H''(Z)}{H(Z)^{6}}\right]\mathrm{d}Z,\label{RT SO 3 2D}
\end{equation}
giving exactly (\ref{Q-P relation SO}).

\subsection{Expression for the dimensionless  pressure drop at the third order}


At the third order, $O(De^{3})$,  (\ref{RT 5 2D}) takes the form 
\begin{eqnarray}
\Delta P_{3} &= &\tilde{\beta}\int_{0}^{H(0)}\left[\mathcal{S}_{zz,2}\hat{U}_{z}\right]_{Z=0}\mathrm{d}Y-\tilde{\beta}\int_{0}^{H(1)}\left[\mathcal{S}_{zz,2}\hat{U}_{z}\right]_{Z=1}\mathrm{d}Y \nonumber\\
  & & +\tilde{\beta}\int_{0}^{1}\int_{0}^{H(Z)}\left(\mathcal{S}_{zz,2}\frac{\partial\hat{U}_{z}}{\partial Z}+\mathcal{S}_{yz,2}\frac{\partial\hat{U}_{z}}{\partial Y}\right)\mathrm{d}Y\mathrm{d}Z,\label{RT TO 1 2D}
\end{eqnarray}where $\mathcal{S}_{zz,2}$ and $\mathcal{S}_{yz,2}$ are given by\begin{subequations}
\begin{equation}
\mathcal{S}_{zz,2}=-U_{z,0}\frac{\partial A_{zz,2}}{\partial Z}-U_{y,0}\frac{\partial A_{zz,2}}{\partial Y}+2\dfrac{\partial U_{z,0}}{\partial Z}A_{zz,2}+2\dfrac{\partial U_{z,0}}{\partial Y}A_{yz,2}+2\dfrac{\partial U_{z,2}}{\partial Y}A_{yz,0},\label{Szz2 2D}
\end{equation}
\begin{eqnarray}
\mathcal{S}_{yz,2}& = &-U_{z,0}\frac{\partial A_{yz,2}}{\partial Z}-U_{z,2}\frac{\partial A_{yz,0}}{\partial Z}-U_{y,0}\frac{\partial A_{yz,2}}{\partial Y}
  \nonumber\\
&& -U_{y,2}\frac{\partial A_{yz,0}}{\partial Y}+\frac{\partial U_{y,0}}{\partial Z}A_{zz,2}+\frac{\partial U_{z,0}}{\partial Y}A_{yy,2}+\frac{\partial U_{z,2}}{\partial Y}A_{yy,0}.
\label{Syz2 2D}
\end{eqnarray}\end{subequations}
Since $\mathcal{S}_{zz,2}$ and $\mathcal{S}_{yz,2}$ depend on the solution from previous orders, we can calculate the third-order pressure drop $\Delta  P_{3}$ using the solution of the Newtonian problem and the solution of the leading-, first- and second-order viscoelastic problems.

The resulting expression for $\Delta  P_{3}$ is
\begin{equation}
\Delta P_{3}=\frac{648\tilde{\beta}(9-\tilde{\beta})}{35}\left(\frac{H'(0)^{2}}{H(0)^{8}}-\frac{H'(1)^{2}}{H(1)^{8}}\right)-\frac{216\tilde{\beta}(8-\tilde{\beta})}{35}\left (\frac{H''(0)}{H(0)^{7}}-\frac{H''(1)}{H(1)^{7}}\right ).\label{RT TO 2 2D}
\end{equation}
In summary, using the reciprocal theorem (\ref{RT 5 2D}) we have determined the dimensionless pressure
drop $\Delta P=\Delta p/(\eta_{0}q\ell/2h_{\ell}^{3})$ as a function
of the shape function $H(Z)$, the viscosity ratio $\tilde{\beta}$ and the
Deborah number $De$ up to $O(De^3)$,
\begin{equation}
\Delta P=\Delta P_{0}+De\Delta P_{1}+De^{2}\Delta P_{2}+De^{3}\Delta P_{3}+O(\epsilon^{2},De^{4}),\label{Pressure drop RT 2D}
\end{equation}
where the expressions for $\Delta P_{0}$, $\Delta P_{1}$, $\Delta P_{2}$, and $\Delta P_{3}$
are given in (\ref{Q-P relation LO}), (\ref{RT FO 3 2D}), (\ref{RT SO 3 2D}), and
(\ref{RT TO 2 2D}), respectively.

\section{Results and comparison with finite-element simulations}\label{Results}

In this section, we present the analytical results for the pressure drop and flow and stress fields of the Oldroyd-B fluid developed in $\mathsection$$\mathsection$  \ref{LA} and \ref{RT section}. We also validate the predictions of our theoretical model by performing 2-D numerical simulations with the finite-element software COMSOL Multiphysics (version 5.6, COMSOL AB, Stockholm, Sweden), with which we compare our analytical results. The details of the numerical procedure are provided in appendix \ref{Appen}.

As an illustrative example, we specifically consider the case of a hyperbolic contracting channel of the form
\begin{equation}
H(Z)=\frac{\alpha}{(\alpha-1)Z+1},\label{Hyperbolic contraction}
\end{equation}
where $\alpha=h_{0}/h_{\ell}$ is a ratio of the heights at the inlet
and outlet; for the contracting geometry we have $\alpha>1$. For the  2-D hyperbolic contracting geometry, (\ref{Hyperbolic contraction}), closed-form analytical  expressions  for the contributions to the pressure drop up to $O(De^3)$ are obtained from (\ref{Q-P relation LO}), (\ref{RT FO 3 2D}), (\ref{RT SO 3 2D}), and
(\ref{RT TO 2 2D}) as
\begin{subequations}
\begin{equation}
\Delta P_{0}=\frac{3}{4}\frac{(1+\alpha)(1+\alpha^{2})}{\alpha^{3}},\label{dP0 2D}
\end{equation}
\begin{equation}
\Delta P_{1}=\frac{9}{2}\tilde{\beta}\frac{1-\alpha^{4}}{\alpha^{4}},\label{dP1 2D}
\end{equation}
\begin{equation}
\Delta P_{2}=\frac{648}{35}\tilde{\beta}\frac{(1+\alpha)(1+\alpha^{2})(1-\alpha)^{2}}{\alpha^{5}},\label{dP2 2D}
\end{equation}
\begin{equation}
\Delta P_{3}=\frac{216}{35}\tilde{\beta}(11-\tilde{\beta})\frac{(1+\alpha)(1+\alpha^{2})(1-\alpha)^{3}}{\alpha^{6}}.\label{dP3 2D}
\end{equation}\label{Hyp_dP}\end{subequations}As expected, (\ref{Hyp_dP}) clearly shows that for the straight channel, $\alpha=1$, the $\Delta P_{1}$, $\Delta P_{2}$, and $\Delta P_{3}$ contributions vanish and the pressure drop of the Oldroyd-B fluid is identical to the pressure drop of the Newtonian fluid with the same zero-shear-rate viscosity.

\subsection{Variation of pressure drop with the Deborah and Weissenberg numbers}

In this work, we mainly present the results for the Oldroyd-B fluid with $\tilde{\beta}=0.4$ in two hyperbolic geometries, which have an identical inlet-to-outlet ratio $\alpha=4$ but different aspect ratios $\epsilon=h_{\ell}/\ell$. The first geometry we consider has $\epsilon=0.02$, for which the assumptions of the lubrication approximation are expected to be well satisfied. 
In addition, aiming to examine the pressure drop in less narrow configurations, the second geometry we study corresponds to $\epsilon=0.1$. For such geometry, even if  $\epsilon=0.1$ can be considered a small parameter, the requirement $\alpha \epsilon \ll 1$, representing the slow variation assumption in the lubrication theory, is not satisfied since $\alpha \epsilon=0.4$ is $O(1)$. However, as we show below, although the lubrication assumptions are not strictly satisfied in this case, our theory captures fairly well the variation of the pressure drop with $De$.





\begin{figure}
 \centerline{\includegraphics[scale=1]{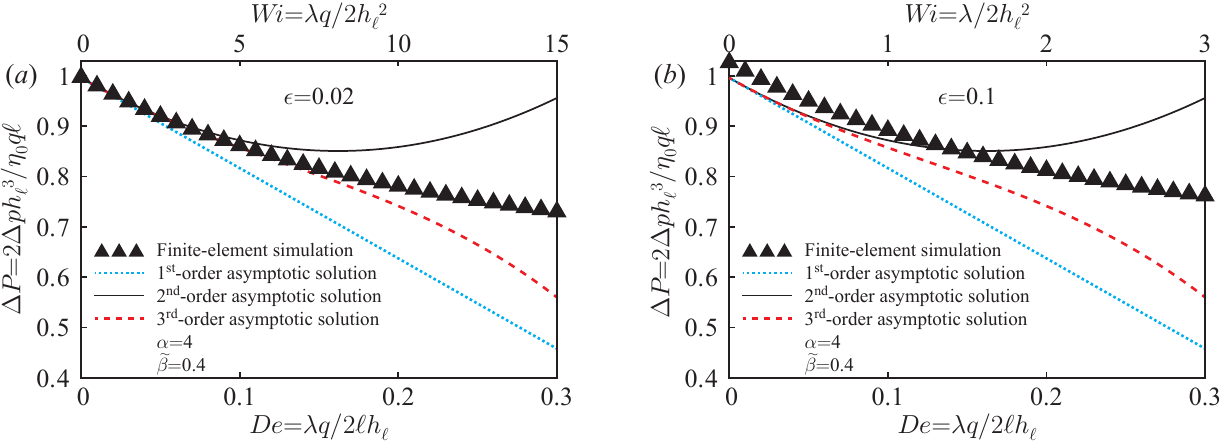}}
\caption{Pressure drop for the Oldroyd-B fluid in a hyperbolic contracting channel described by equation~(\ref{Hyperbolic contraction}). ($a,b$) Dimensionless pressure drop $\Delta P=\Delta p/(\eta_{0}q\ell/2h_{\ell}^{3})$ as a function of $De=\lambda q/(2\ell h_{\ell})$ (or $Wi=\lambda q/(2h_{\ell}^{2}))$ for $\epsilon=0.02$ ($a$) and $\epsilon=0.1$ ($b$). Black triangles ($\blacktriangle$) represent the results of the finite-element simulation. Dotted cyan (\textcolor{cyan}{$\cdot$$\cdot$$\cdot$$\cdot$}) lines represent the first-order asymptotic solution, given by (\ref{dP0 2D})$-$(\ref{dP1 2D}). Black solid (\textcolor{black}{---}) lines represent the second-order asymptotic solution, given by (\ref{dP0 2D})$-$(\ref{dP2 2D}). Red dashed (\textcolor{red}{- -}) lines represent the third-order asymptotic solution, given by (\ref{dP0 2D})$-$(\ref{dP3 2D}). All calculations were performed using $\alpha=4$ and $\tilde{\beta}=0.4$.}\label{F2}
\end{figure}
We present the non-dimensional pressure drop $\Delta P=\Delta p/(\eta_{0}q\ell/2h_{\ell}^{3})$ as a function of $De=\lambda q/(2\ell h_{\ell})$ (or $Wi=\lambda q/(2h_{\ell}^{2}))$ in figure \ref{F2}($a,b$) for the Oldroyd-B fluid in a hyperbolic contracting channel for $\epsilon=0.02$ ($a$) and $\epsilon=0.1$ ($b$), with $\alpha=4$ and $\tilde{\beta}=0.4$. Cyan dotted lines represent the first-order asymptotic solution, given by (\ref{dP0 2D})$-$(\ref{dP1 2D}), black solid lines represent the second-order asymptotic solution, given by (\ref{dP0 2D})$-$(\ref{dP2 2D}), and red dashed lines represent the third-order asymptotic solution, given by (\ref{dP0 2D})$-$(\ref{dP3 2D}). Black triangles represent the results of the numerical simulation obtained from calculating the pressure drop along the centerline ($Y=0$). We note that while our analysis assumes $De\ll1$, where Deborah number is the product of the relaxation time and the characteristic extensional rate of the flow, the Weissenberg number $Wi$, which is the product of the relaxation time and the characteristic shear rate of the flow, is $O(1)$ \citep[see also][]{ahmed2021new}. To further highlight this point, we present our results both as a function of $De$ and $Wi$.

Similar to previous numerical reports using the Oldroyd-B model for studying the flow of Boger fluids in 2-D abruptly contracting geometries \citep[see, e.g.,][]{aboubacar2002highly,alves2003benchmark,binding2006contraction,aguayo2008excess}, our high-order analytical and numerical simulations in figure \ref{F2}($a,b$) predict a monotonic decrease in the pressure drop with increasing $De$ (or $Wi$). 
In addition, the results in figure \ref{F2}($a$)
clearly show that accounting for higher orders of the analytical solutions for the pressure drop significantly improves the agreement with the numerical simulation results for  $\epsilon=0.02$, yielding 
a relative error of $\approx 5$ $\%$ for up to $De=0.2$, corresponding to $Wi=10$. 
For the case of $\epsilon=0.1$, shown in figure \ref{F2}($b$), the third-order asymptotic solution, given by (\ref{dP0 2D})$-$(\ref{dP3 2D}), slightly underpredicts the pressure drop, yet even for $De=0.2$, corresponding to $Wi=2$, it results in a modest relative error of $\approx 9$ $\%$. 


\subsection{Assessing the effect of different contributions to the pressure drop}\label{diff_con}

The results presented in the previous subsection predict a reduction in the pressure drop with increasing $De$ or $Wi$ for the Oldroyd-B fluid in a hyperbolic contracting channel. In this subsection, to provide insight into the source of such a reduction, we elucidate the relative importance of different contributions to the pressure drop using our analytical predictions and numerical simulations.
To this end, we integrate the momentum equation (\ref{Momentum z ND}) with respect to $Z$ from 0 to 1 along the centerline ($Y=0$) and obtain the dimensionless pressure drop,
\begin{eqnarray}
\Delta P&=&\underset{\mbox{\ding{172}}}{\underbrace{\epsilon^{2}\left[\left.\frac{\partial U_{z}}{\partial Z}\right|_{Y=0,\,Z=0}-\left.\frac{\partial U_{z}}{\partial Z}\right|_{Y=0,\,Z=1}\right]}}+\underset{\mbox{\ding{173}}}{\underbrace{\int_{1}^{0}\left.\frac{\partial^{2}U_{z}}{\partial Y^{2}}\right|_{Y=0}\mathrm{d}Z}} \nonumber\\
  &  &\mbox{\hspace{5mm}} + \underset{\mbox{\ding{174}}}{\underbrace{\tilde{\beta}De\Delta\mathcal{S}_{zz}}}+\underset{\mbox{\ding{175}}}{\underbrace{\tilde{\beta}De\int_{1}^{0}\left.\frac{\partial\mathcal{S}_{yz}}{\partial Y}\right|_{Y=0}\mathrm{d}Z}},\label{dP 2D diffrent contr.}
\end{eqnarray}
where $\Delta P=P(0,0)-P(0,1)$ and $\Delta\mathcal{S}_{zz}=\mathcal{S}_{zz}(0,0)-\mathcal{S}_{zz}(0,1)$. We note that for a general geometry the axial pressure drop may strongly depend on the $Y$ coordinate along which it is evaluated. However, since for narrow geometries $P=P(Z)+O(\epsilon^{2})$, i.e., the pressure is independent of $Y$ up to $O(\epsilon^{2})$, we expect the results to be weakly dependent on the value of $Y$ along which the integration over $Z$ is performed.
\begin{figure}
 \centerline{\includegraphics[scale=1]{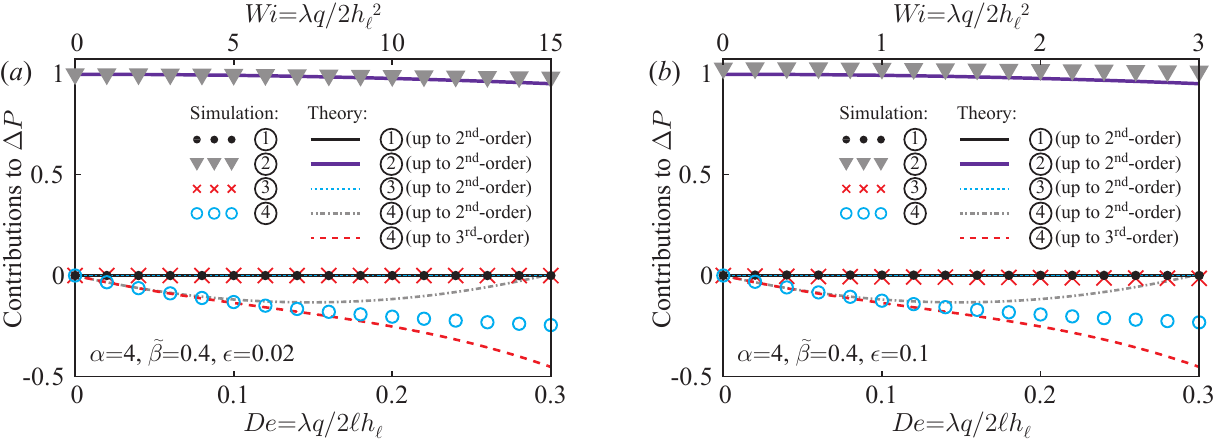}}
\caption{Contributions to the pressure drop of the Oldroyd-B fluid in a hyperbolic contracting channel. ($a,b$) Different contributions to the pressure drop as a function of $De=\lambda q/(2\ell h_{\ell})$ (or $Wi=\lambda q/(2h_{\ell}^{2}))$ for $\epsilon=0.02$ ($a$) and $\epsilon=0.1$ ($b$). Dots, triangles, crosses, and circles represent the  \ding{172}$-$\ding{175} contributions extracted from 2-D numerical simulations. Black solid, purple solid, cyan dotted, and gray dashed-dot lines represent the  \ding{172}$-$\ding{175} contributions obtained from the asymptotic solution up to $O(De^2$).
Red dashed lines represent the analytically obtained \ding{175} contribution up to $O(De^3)$. All calculations were performed using $\alpha=4$ and $\tilde{\beta}=0.4$.}\label{F3}
\end{figure}

Equation (\ref{dP 2D diffrent contr.}) clearly shows that the pressure drop of viscoelastic flow of consists of four contributions. The first (\ding{172}) and third (\ding{174}) terms on the right-hand
side of (\ref{dP 2D diffrent contr.}) represent the contribution of the Newtonian  and  viscoelastic viscous axial stress differences, respectively. The second (\ding{173}) and fourth (\ding{175}) terms represent, respectively, the contribution of the Newtonian and viscoelastic viscous shear stresses. Since the first term, \ding{172}, scales as $O(\epsilon^2)$, we expect it to be negligible for narrow geometries under consideration.

The different contributions to the pressure drop as a function of $De$ (or $Wi$) are shown in figure \ref{F3}($a,b$) for $\epsilon=0.02$ ($a$) and $\epsilon=0.1$ ($b$), with $\alpha=4$ and $\tilde{\beta}=0.4$. Dots, triangles, crosses, and circles, respectively, represent the  contributions \ding{172}$-$\ding{175}  extracted from numerical simulations. Black solid, purple solid, cyan dotted, and gray dashed-dot lines represent the  \ding{172}$-$\ding{175} contributions obtained from the asymptotic solution up to $O(De^2$).
Red dashed lines represent the analytically obtained contribution \ding{175}  up to $O(De^3)$.


First, as expected, both our analytical and numerical simulations show that the \ding{172} term has a negligible contribution to the pressure drop. Second, somewhat surprisingly, from figure \ref{F3}($a,b$) it follows that the \ding{174} term, representing the viscoelastic viscous axial stress difference along the centerline, also has a negligible contribution to the pressure drop.
We rationalize the latter by noting that, for convenience, we have calculated the pressure drop along the centerline, and for $Y=0$,  the viscoelastic viscous axial stress difference is indeed negligible. However, since the pressure is independent of $Y$ to $O(\epsilon^{2})$, (\ref{Momentum z ND}) can be integrated with respect to $Z$ from 0 to 1, while setting the different values of $Y$, for which the viscoelastic viscous axial stress difference may have an apparent contribution to the pressure drop (see figure \ref{F4}). 

It is evident from figure \ref{F3}($a,b$) that only the \ding{173} and \ding{175} terms, which are associated with the Newtonian and viscoelastic viscous shear stresses, have a significant contribution to the pressure drop, calculated along $Y=0$. The \ding{173} term shows only a weak dependence on $De$ and has approximately a constant value, corresponding to the Newtonian case. Such a weak dependence on $De$ is expected, since \ding{173} is related to the velocity $U_{z}=U_{z,0}+De^2U_{z,2}+O(De^3,\epsilon^2)$, which has only $O(De^2)$ and higher contributions. We observe an excellent agreement between our analytical predictions and the results of the numerical simulations for \ding{173} in the case $\epsilon=0.02$ throughout the investigated range of $De$ number. Moreover, although for $\epsilon=0.1$ the lubrication assumptions are not strictly satisfied, our asymptotic solution for \ding{173} is in fair agreement with numerical simulations for this case as well.
We note that a small discrepancy exists even for the Newtonian case ($De=0$), thus indicating that the error is due to the non-fulfillment of the lubrication assumptions rather than the low-$De$ analysis.

Unlike \ding{173}, the \ding{175} term strongly depends on $De$, and both our third-order asymptotic solution (red dashed lines) and numerical simulations (triangles) predict a monotonic decrease with $De$, which is the main source of reduction in the pressure drop observed in figure \ref{F2}. 
We, therefore, may conclude that for narrow configurations, such as those shown in figure \ref{F1}, the pressure drop, calculated along $Y=0$, is determined from the balance between the  \ding{173} and \ding{175} terms and the reduction in the pressure drop for contracting channels is due to the viscoelastic viscous shear stress term \ding{175}.

To further highlight the latter point, let us calculate the first-order correction to the pressure drop, $\Delta P_{1}$, using (\ref{dP 2D diffrent contr.}). As there are no velocity components at the first order (see  $\mathsection$ \ref{FO section}) and the third term $\tilde{\beta}\Delta\mathcal{S}_{zz,0}$ vanishes at the centerline, we obtain that $\Delta P_{1}$ can be written as
\begin{equation}
    \Delta P_{1}=\tilde{\beta}\int_{1}^{0}\left.\frac{\partial\mathcal{S}_{yz,0}}{\partial Y}\right|_{Y=0}\mathrm{d}Z.\label{dP1 diff. contr.}
\end{equation}
Using (\ref{Continuity ND 2D LO}), (\ref{Ayy ND 2D LO}), (\ref{Ayz ND 2D LO}), and (\ref{Syz0 2D}), the argument $\partial\mathcal{S}_{yz,0}/\partial Y|_{Y=0}$ can be expressed as
\begin{eqnarray}
\left.\frac{\partial\mathcal{S}_{yz,0}}{\partial Y}\right|_{Y=0} &= &-\left[U_{z,0}\dfrac{\partial^{3}U_{z,0}}{\partial Y^{2}\partial Z}+\dfrac{\partial U_{y,0}}{\partial Y}\dfrac{\partial^{2}U_{z,0}}{\partial Y^{2}}+2\dfrac{\partial^{2}U_{z,0}}{\partial Y^{2}}\dfrac{\partial U_{z,0}}{\partial Z}\right]_{Y=0} \nonumber\\
  & =& -\left[U_{z,0}\dfrac{\partial^{3}U_{z,0}}{\partial Y^{2}\partial Z}-\dfrac{\partial U_{z,0}}{\partial Z}\dfrac{\partial^{2}U_{z,0}}{\partial Y^{2}}+2\dfrac{\partial^{2}U_{z,0}}{\partial Y^{2}}\dfrac{\partial U_{z,0}}{\partial Z}\right]_{Y=0}\nonumber\\
    & =& -\left[U_{z,0}\dfrac{\partial^{3}U_{z,0}}{\partial Y^{2}\partial Z}+\dfrac{\partial^{2}U_{z,0}}{\partial Y^{2}}\dfrac{\partial U_{z,0}}{\partial Z}\right]_{Y=0}=-\left[\frac{\partial}{\partial Z}\left(U_{z,0}\dfrac{\partial^{2}U_{z,0}}{\partial Y^{2}}\right)\right]_{Y=0}\nonumber\\
    & =& -\left[\frac{\partial^{2}}{\partial Z\partial Y}\left(U_{z,0}\dfrac{\partial U_{z,0}}{\partial Y}\right)\right]_{Y=0}=-\left[\frac{\partial}{\partial Z}\left(U_{z,0}\frac{\partial A_{yz,0}}{\partial Y}\right)\right]_{Y=0},\label{Syx0 der.}
\end{eqnarray}where we further used the fact that both $U_{y,0}$ and $\partial U_{z,0}/\partial Y$ vanish at the centerline. 
Substituting (\ref{Syx0 der.}) into (\ref{dP1 diff. contr.}), and using (\ref{U(X,Y) LO e}) and (\ref{Axx, Axy LO e}), yields  
\begin{equation}
\Delta P_{1}=-\tilde{\beta}\left[\left(U_{z,0}\frac{\partial A_{yz,0}}{\partial Y}\right)_{Y=0,Z=0}-\left(U_{z,0}\frac{\partial A_{yz,0}}{\partial Y}\right)_{Y=0,Z=1}\right]=\frac{9\tilde{\beta}}{2}\left(\frac{1}{H(0)^{4}}-\frac{1}{H(1)^{4}}\right),\label{dP 2D diffrent contr.-1-2-1}
\end{equation}clearly showing that the first-order pressure drop, calculated along the centreline, arises due to the velocity variation and viscoelastic shear stresses gradients.

\subsection{Comparison between the analytical predictions and the 2-D numerical simulations for the axial velocity and polymer stress contributions}
\begin{figure}
 \centerline{\includegraphics[scale=1]{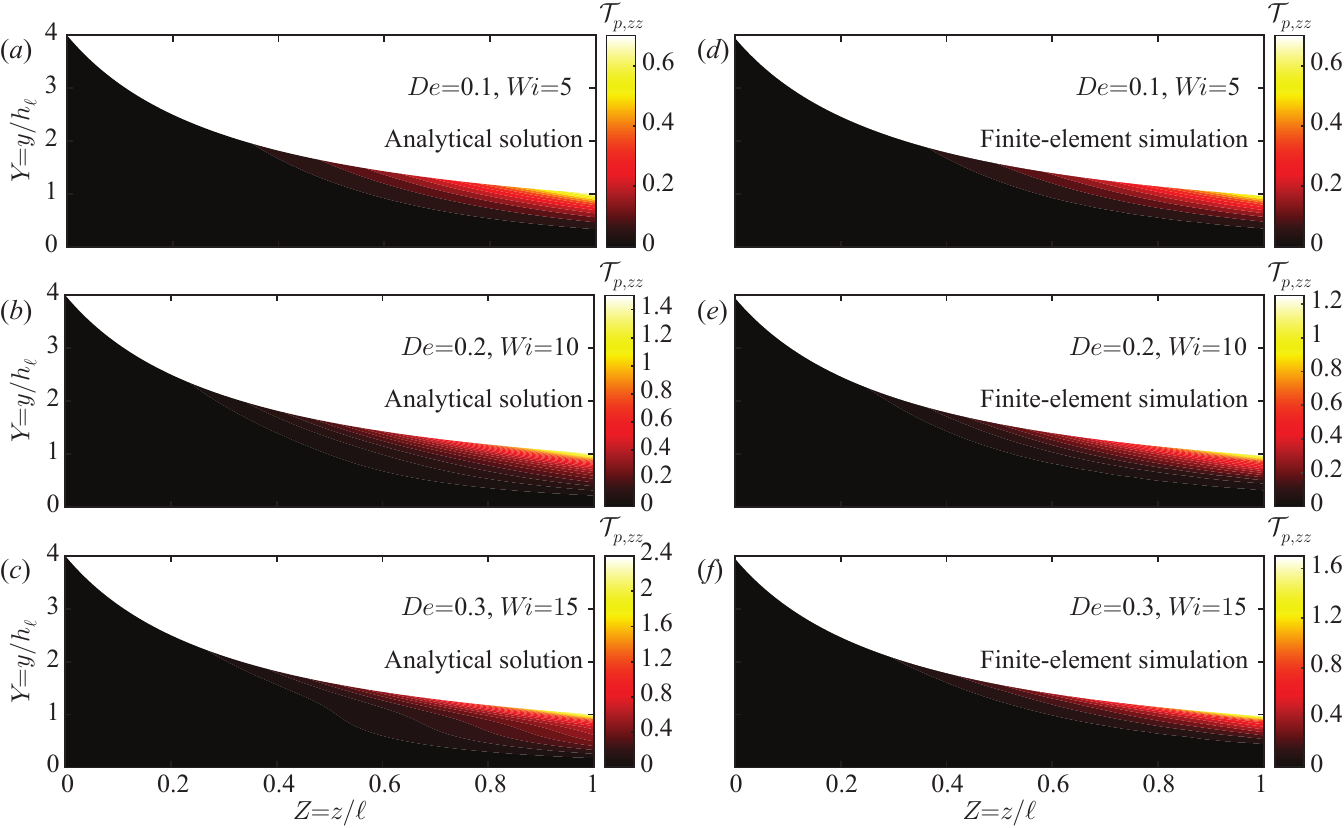}}
\caption{Contour plot of the axial polymer stress distribution, $\mathcal{T}_{p,zz}$, as a function of the $(Y,Z)$ coordinates for $De=0.1$ ($Wi=5$) ($a,d$), $De=0.2$ ($Wi=10$) ($b,e$) and $De=0.3$ ($Wi=15$) ($c,f$), obtained from our analytical  theory ($a$--$c$) and 2-D numerical simulations ($d$--$f$). All calculations were performed using $\epsilon=0.02$, $\alpha=4$, and $\tilde{\beta}=0.4$.}\label{F4}
\end{figure}
\begin{figure}
 \centerline{\includegraphics[scale=1]{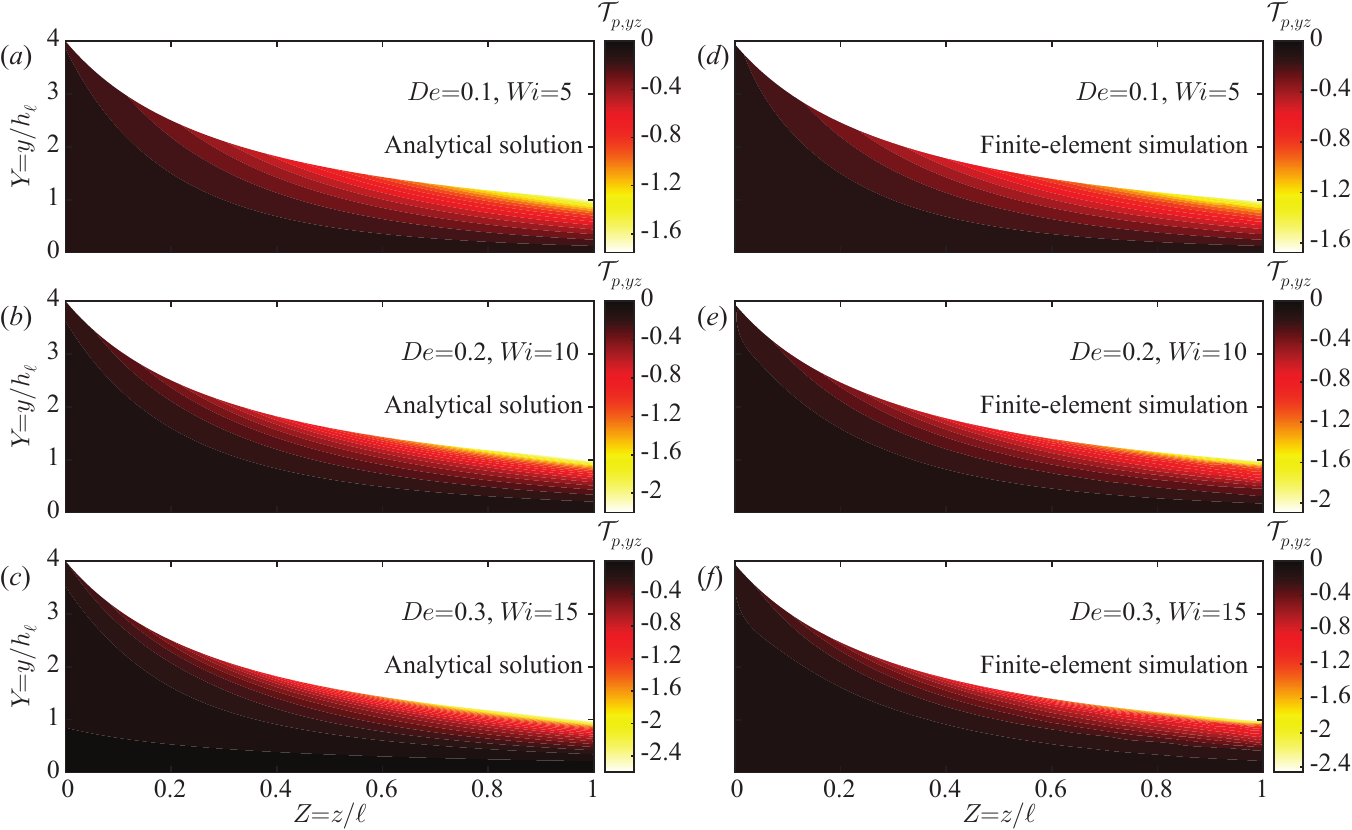}}
\caption{Contour plot of the polymer shear stress distribution, $\mathcal{T}_{p,yz}$, as a function of the $(Y,Z)$ coordinates for $De=0.1$ ($Wi=5$) ($a,d$), $De=0.2$ ($Wi=10$) ($b,e$) and $De=0.3$ ($Wi=15$) ($c,f$), obtained from our analytical  theory ($a$--$c$) and 2-D numerical simulations ($d$--$f$). All calculations were performed using $\epsilon=0.02$, $\alpha=4$, and $\tilde{\beta}=0.4$.}\label{F5}
\end{figure}

The theoretical results derived in $\mathsection$ \ref{LA} allow determination of closed-form analytical expressions for the velocity and pressure, as well as solvent and polymer stress distributions, which can then be compared with the results of numerical simulations. It is of particular interest to compare the results for the polymer stress distribution, especially the  $\mathcal{T}_{p,zz}$ and $\mathcal{T}_{p,yz}$ components, whose gradients contribute to the axial pressure gradient, ultimately resulting in the pressure drop. Due to symmetry along the $Y=0$, below we show the polymer stress and velocity distributions only in the half domain, $Y\geq0$.

We present in figures \ref{F4} and \ref{F5} a comparison of our analytical predictions ($a$--$c$) and finite-element simulation results ($d$--$f$) for the axial and shear polymer stress distribution, $\mathcal{T}_{p,zz}$ and $\mathcal{T}_{p,yz}$, respectively, for different values of $De$, with $\epsilon=0.02$, $\alpha=4$, and $\tilde{\beta}=0.4$. Clearly, for $De=0.1$ and $De=0.2$ there is good agreement between our analytical predictions and the numerical results for both $\mathcal{T}_{p,zz}$ and $\mathcal{T}_{p,yz}$. However, as expected, when $De$ increases, the agreement deteriorates, and for $De=0.3$, our analytical solution overpredicts the magnitude of $\mathcal{T}_{p,zz}$ and $\mathcal{T}_{p,yz}$ on the wall and does not capture exactly the polymer stress distribution in the entire domain. As $\mathcal{T}_{p,yz}$ and $\mathcal{S}_{yz}$ are related through (\ref{T-S relation}), the latter observation is consistent with the discrepancy between theory and simulations  observed for $De=0.3$ in figure \ref{F3}($a$) for the pressure drop contribution related to the $\partial\mathcal{S}_{yz}/\partial Y$ (term \ding{175}).


It is evident from figures \ref{F4} and \ref{F5} that $\mathcal{T}_{p,zz}$ is positive and $\mathcal{T}_{p,yz}$ is negative for $Y>0$, with a minimum magnitude on the centreline and a maximum magnitude on the wall at the outlet, i.e., $(Y, Z)=(1,1)$. These results and the $\mathcal{T}_{p,zz}$ and $\mathcal{T}_{p,yz}$ distributions are in qualitative agreement with the numerical results of \citet{nystrom2016extracting} for the axial and shear polymer stress contributions of a viscoelastic fluid, described by the FENE-CR model, in an axisymmetric contracting hyperbolic channel, shown in their figure 7($b,c$).
It is also worth noting that all of the presented analytical and numerical results here for $\mathcal{T}_{p,zz}$ and $\mathcal{T}_{p,yz}$ are $O(1)$, thus clearly showing that our scalings in (\ref{Non-dimensional variables 3}) and (\ref{Non-dimensional variables 4}) for narrow geometries are representative, consistent with the studies on thin films and lubrication problems \citep{tichy1996non,zhang2002surfactant,saprykin2007free,ahmed2021new}.  We note that in most studies all the components of the polymer stress tensor were scaled with the same scaling $\eta_{0}u_{c}/h_{\ell}$. While such a scaling holds for geometries with $h_{\ell}/\ell=O(1)$, for narrow geometries it becomes inappropriate.

\begin{figure}
 \centerline{\includegraphics[scale=1]{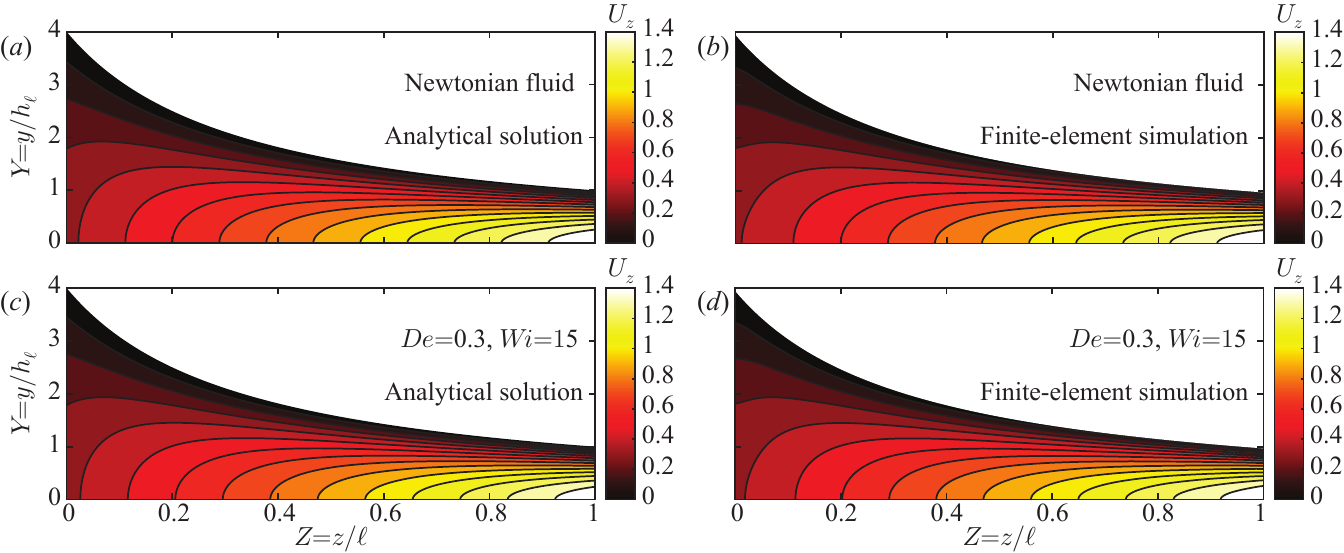}}
\caption{Comparison of analytical predictions ($a,c$) and finite-element simulation results ($b,d$) for contours of the axial velocity, $U_{z}$, as a function of the $(Y,Z)$ coordinates for the Newtonian fluid ($a,b$) and Oldroyd-B fluid ($c,d$) with $De=0.3$ ($Wi=15$) in the case of a hyperbolic contracting channel. All calculations were performed using $\epsilon=0.02$, $\alpha=4$, and $\tilde{\beta}=0.4$.}\label{F6}
\end{figure}

In addition to the polymer stress distribution, we compare our analytical predictions for the axial velocity with the results of the numerical simulations. Figure \ref{F6} shows a comparison of analytical predictions ($a,c$) and finite-element simulation results ($b,d$) for contours of the axial velocity, $U_{z}$, as a function of the $(Y,Z)$ coordinates for the Newtonian fluid ($a,b$) and Oldroyd-B fluid ($c,d$) with $De=0.3$ ($Wi=15$), $\epsilon=0.02$, $\alpha=4$, and $\tilde{\beta}=0.4$. First, we observe excellent agreement between the analytical and numerical results for the axial velocity $U_{z}$ for both $De=0$ (Newtonian case) and $De=0.3$. Second and more importantly, the axial velocity distribution for $De=0.3$ seems nearly identical to the Newtonian case, which might seem surprising given the observed pressure drop reduction for $De=0.3$. As we noted in $\mathsection$ \ref{diff_con}, the reason for this behavior is the weak dependence of the velocity on the Deborah number, so that the viscoelastic effects start to affect the flow field only at $O(De^2$) and higher orders. 

 \subsection{Effect of the inlet-to-outlet aspect ratio and polymer-to-solvent viscosity ratio}

\begin{figure}
 \centerline{\includegraphics[scale=1]{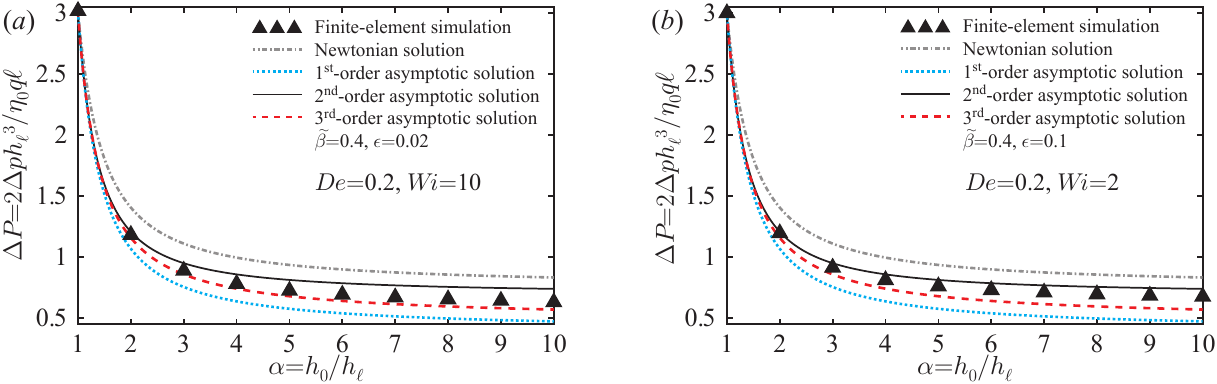}}
\caption{The effect of the inlet-to-outlet ratio $\alpha=h_{0}/h_{\ell}$ on the pressure drop of the Oldroyd-B fluid in a hyperbolic contracting channel. ($a$) Dimensionless pressure drop $\Delta P$ as a function of $\alpha$ for $\epsilon=0.02$ and $De=0.2$ ($Wi=10$). ($b$) Dimensionless pressure drop $\Delta P$ as a function of $\alpha$ for $\epsilon=0.1$ and $De=0.2$ ($Wi=2$). Black triangles ($\blacktriangle$) represent the results of the finite-element simulations. Gray dashed-dot (\textcolor{gray}{{-} $\cdot$ {-}}) lines represent the Newtonian solution, given by (\ref{dP0 2D}). Cyan dotted (\textcolor{cyan}{$\cdot$$\cdot$$\cdot$$\cdot$}) lines represent the first-order asymptotic solution, given by (\ref{dP0 2D})$-$(\ref{dP1 2D}). Black solid (\textcolor{black}{---}) lines represent the second-order asymptotic solution, given by (\ref{dP0 2D})$-$(\ref{dP2 2D}). Red dashed (\textcolor{red}{- -}) lines represent the third-order asymptotic solution, given by (\ref{dP0 2D})$-$(\ref{dP3 2D}). All calculations were performed using $\tilde{\beta}=0.4$.}\label{F7}
\end{figure}

 In this section, we explore the effect of the inlet-to-outlet aspect ratio $\alpha=h_{0}/h_{\ell}$  and polymer-to-solvent viscosity ratio $\eta_p/\eta_{s}$ on the pressure drop. 
First, in 
figure \ref{F7}($a,b$)  we present the non-dimensional pressure drop $\Delta P=\Delta p/(\eta_{0}q\ell/2h_{\ell}^{3})$ as a function of $\alpha$ for $\epsilon=0.02$ ($a$) and $\epsilon=0.1$ ($b$), with $De=0.2$ and $\tilde{\beta}=0.4$.  Gray dashed-dot lines represent the leading-order (Newtonian) asymptotic solution, given by (\ref{dP0 2D}), and as earlier,
cyan dotted lines represent the first-order asymptotic solution, given by (\ref{dP0 2D})$-$(\ref{dP1 2D}), black solid lines represent the second-order asymptotic solution, given by (\ref{dP0 2D})$-$(\ref{dP2 2D}), and red dashed lines represent the third-order asymptotic solution, given by (\ref{dP0 2D})$-$(\ref{dP3 2D}). Black triangles represent the results of the numerical simulations.
As expected, when increasing $\alpha=h_{0}/h_{\ell}$ (or $h_{0}$), while fixing the values of $h_{\ell}$ and $q$, the $\Delta P=\Delta p/(\eta_{0}q\ell/2h_{\ell}^{3})$, which can be viewed as the dimensionless  hydrodynamic resistance,  monotonically decreases. Moreover, for a given value of  $\alpha=h_{0}/h_{\ell}$,  the pressure drop of the Oldroyd-B fluid is smaller than that of a Newtonian fluid, consistent with the results of figure \ref{F2}.

For small values of $\alpha$, figure \ref{F7}($a,b$) shows good agreement between the third-order asymptotic solution and numerical simulation results for both $\epsilon=0.02$ and $\epsilon=0.1$. However, as  $\alpha$ increases, the agreement between the theory and simulations deteriorates because the assumptions of the lubrication theory become less well satisfied. Nevertheless,  even the case of $\alpha=10$ results in relative errors of only $\approx10$ $\%$ and $\approx16$ $\%$ for $\epsilon=0.02$ and $\epsilon=0.1$, respectively. The latter result for $\epsilon=h_{\ell}/\ell=0.1$ and  $\alpha \epsilon=h_{0}/\ell=1$ clearly indicates that our theory is applicable not only to  narrow geometries but also can be used for geometries with a high aspect ratio,  and can still reasonably predict the pressure drop.

\begin{figure}
 \centerline{\includegraphics[scale=1]{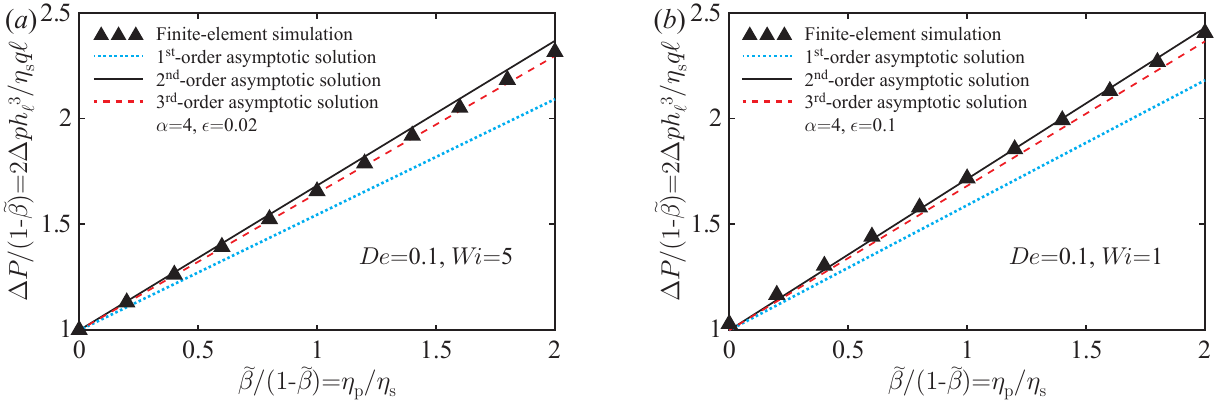}}
\caption{The effect of the polymer contribution to the viscosity on the pressure drop of the Oldroyd-B fluid in a hyperbolic contracting channel. ($a$) Pressure drop $\Delta p$ scaled by $\eta_{s}q\ell/2h_{\ell}^{3}$
as a function of the polymer to solvent viscosity ratio $\eta_p/\eta_{s}$ for $\epsilon=0.02$ and $De=0.1$ ($Wi=5$). ($b$) Pressure drop $\Delta p$ scaled by $\eta_{s}q\ell/2h_{\ell}^{3}$
as a function of the polymer to solvent viscosity ratio $\eta_p/\eta_{s}$ for $\epsilon=0.1$ and $De=0.1$ ($Wi=1$). Black triangles ($\blacktriangle$) represent the results of the finite-element simulation. Cyan dotted (\textcolor{cyan}{$\cdot$$\cdot$$\cdot$$\cdot$}) lines represent the first-order asymptotic solution, given by (\ref{dP0 2D})$-$(\ref{dP1 2D}). Black solid (\textcolor{black}{---}) lines represent the second-order asymptotic solution, given by (\ref{dP0 2D})$-$(\ref{dP2 2D}). Red dashed (\textcolor{red}{- -}) lines represent the third-order asymptotic solution, given by (\ref{dP0 2D})$-$(\ref{dP3 2D}). All calculations were performed using $\alpha=4$.}\label{F8}
\end{figure}

Finally, we consider the effect of the polymer-to-solvent viscosity ratio $\eta_p/\eta_{s}$ on the pressure drop in figure \ref{F8}($a,b$) for $\epsilon=0.02$ ($a$) and $\epsilon=0.1$ ($b$), with $De=0.1$ and $\alpha=4$.
We note that it is of more practical interest to discuss the effect of $\eta_p/\eta_{s}$ rather than $\tilde{\beta}=\eta_{p}/\eta_{0}$, since typically in the experiments the viscosity of the solvent $\eta_{s}$ remains fixed, while the polymer viscosity $\eta_{p}$ may change through modifying the polymer concentration, and thus the total viscosity $\eta_{0}=\eta_{s}+\eta_{p}$ may also vary. Now, as $\eta_{0}$ varies, we find it is more appropriate to present in figure \ref{F8}($a,b$) the pressure drop $\Delta p$ scaled by $\eta_{s}q\ell/2h_{\ell}^{3}$, which is $\Delta P/(1-\tilde{\beta})$, rather than $\Delta P$. It is evident from figure \ref{F8}($a,b$) that for a given value of $De$ (or $Wi$), the pressure drop increases linearly with $\eta_p/\eta_{s}$. This
behaviour can be explained using (\ref{dP 2D diffrent contr.}) and noting that the first and third terms have negligible contribution to the pressure drop, as shown in figure \ref{F3}, so that $\Delta P/(1-\tilde{\beta})$ is approximately:
\begin{equation}
 \frac{\Delta P}{1-\tilde{\beta}}\approx\underset{\mbox{\ding{173}}}{\underbrace{\left(1+\frac{\eta_{p}}{\eta_{s}}\right)\int_{1}^{0}\left.\frac{\partial^{2}U_{z}}{\partial Y^{2}}\right|_{Y=0}\mathrm{d}Z}} 
  +\underset{\mbox{\ding{175}}}{\underbrace{\frac{\eta_{p}}{\eta_{s}}De\int_{1}^{0}\left.\frac{\partial\mathcal{S}_{yz}}{\partial Y}\right|_{Y=0}\mathrm{d}Z}},\label{dP 2D diffrent contr. polymer}   
\end{equation}
clearly showing that $\Delta P/(1-\tilde{\beta})$ scales linearly with $\eta_p/\eta_{s}$.

For $\epsilon=0.02$, we observe a good agreement between the third-order asymptotic solution (red dashed line) and numerical simulation results (black triangles). For  $\epsilon=0.1$, however, the third-order asymptotic solution slightly underpredicts the numerically obtained pressure drop, consistent with our previous results shown in figure \ref{F2}($b$). Nevertheless, the resulting relative error is below 5 $\%$ throughout the investigated range of parameters.

\section{Concluding remarks}\label{CR}

In this work, we studied the pressure-driven flow of an Oldroyd-B fluid in arbitrarily shaped, narrow channels and developed a theoretical framework for calculating the velocity and stress fields and the $q-\Delta p$ relation. Using the lubrication approximation, we first identified the appropriate characteristic scales and dimensionless parameters governing the viscoelastic flow in narrow geometries. We then employed a perturbation expansion in powers of $De$ and provided analytical expressions for the velocity and stress fields and the flow rate$-$pressure drop relation in the weakly viscoelastic limit up to $O(De^2)$. We further exploited the reciprocal theorem to obtain the $q-\Delta p$ relation at the next order, $O(De^3)$, using only the velocity and stress fields at the previous orders, eliminating the need to solve the viscoelastic flow problem at $O(De^3)$.

To validate the results of our theoretical model, we performed 2-D numerical simulations of the viscoelastic flow, described by the Oldroyd-B model, in a hyperbolic, symmetric contracting channel for the flow-rate-controlled situation. For geometries that satisfy well the lubrication assumptions, we found excellent agreement between the velocity, polymer stress, and pressure drop predicted by our theory and those obtained from the numerical simulations. Furthermore, we showed that our theory is applicable not only to narrow geometries but it also can be used for geometries with a high aspect ratio, while still reasonably predicting the pressure drop in the weakly viscoelastic limit. 

Both our theory and simulations showed a weak dependence of the velocity field of an Oldroyd-B fluid on the Deborah number so that it can be approximated as Newtonian. In contrast, we demonstrated that the pressure drop of an Oldroyd-B fluid strongly depends on the viscoelastic effects and monotonically decreases with increasing $De$, similar to previous numerical reports on 2-D abruptly contracting geometries \citep{aboubacar2002highly,alves2003benchmark,binding2006contraction,aguayo2008excess}.
 To understand the cause for such pressure drop reduction, which has been largely unexplored to date, we elucidated the relative importance of different terms  contributing to the pressure drop along the symmetry line (see (\ref{dP 2D diffrent contr.})).  We identified that a pressure drop reduction for narrow contracting geometries is primarily due to viscoelastic shear stresses gradients (term \ding{175} in (\ref{dP 2D diffrent contr.})), while viscoelastic axial stresses (term \ding{174} in (\ref{dP 2D diffrent contr.})) make a negligible  contribution to  the pressure drop, calculated along the symmetry line. 
 
Our theoretical approach is not restricted to the case of two-dimensional channels and can be utilized to calculate the flow rate$-$pressure drop relation in narrow axisymmetric geometries. We also expect our results to directly apply to narrow and shallow three-dimensional channels of length $\ell$, width
$w$ and height $h$, where $h\ll w\ll\ell$, or  $h/\ell\ll 1$ and $h/w\ll 1$. Nevertheless, further investigation would be required to assess the range of validity of this approximation. 

One interesting extension of the present work, which relies on the leading-order lubrication theory, is to calculate the high-order perturbative corrections to the pressure drop, following, e.g., \citet{tavakol2017extended}. We anticipate that with such corrections, our framework will allow accurate prediction of the pressure drop in geometries with a modest ratio. Finally, while we considered the Oldroyd-B model to describe the viscoelasticity and predicted a monotonic reduction in the pressure drop with increasing $De$, as a future research direction, it is interesting to analyze more complex constitutive models that incorporate additional microscopic features of polymer solutions and to study the effect of these features on the pressure drop. 
Although a more complex model may pose significant challenges for analytical progress, we anticipate the theoretical framework presented here may still allow the development of a simplified, reduced-order model, amenable to asymptotic/numerical investigations.

\backsection[Acknowledgements]{We thank C. A. Browne, S. S. Datta, and L. G. Leal for helpful discussions. We thank D. Ilssar and D. M. Kochmann for providing us computational facilities for performing numerical simulations.}

\backsection[Funding]{This research was partially supported by NSF through the Princeton University’s Materials Research Science and Engineering Center DMR-2011750. E. B. acknowledges the support of the Yad Hanadiv (Rothschild) Foundation and the Zuckerman STEM Leadership Program.}

\backsection[Declaration of interests]{The authors report no conflict of interest.}


\backsection[Author ORCIDs]{
\\Evgeniy Boyko https://orcid.org/0000-0002-9202-5154;\\
Howard A. Stone https://orcid.org/0000-0002-9670-0639.}



\appendix

\section{Details of numerical simulations}\label{Appen}

\begin{figure}
 \centerline{\includegraphics[scale=1.07]{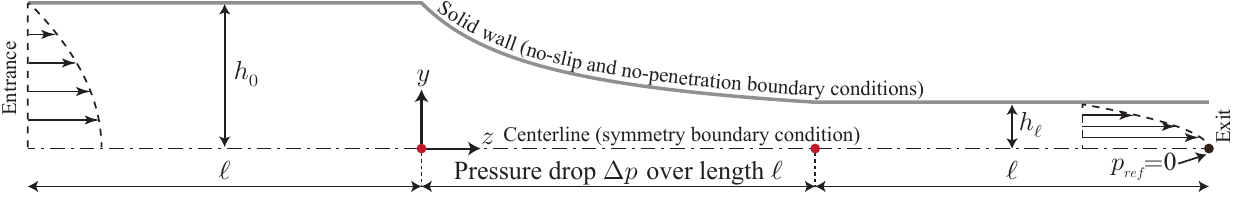}}
\caption{Schematic illustration of the two-dimensional hyperbolic contracting channel with straight entrance and exit regions of length $\ell$  used in finite-element numerical simulations for the pressure-driven flow of the Oldroyd-B fluid. }\label{F9}
\end{figure}
\begin{table}
  \begin{center}
\def~{\hphantom{0}}
  \begin{tabular}{cccccccccc}
    $\ell$  (mm)  &  $\eta_{0}$ (Pa s)&  $\eta_{s}$ (Pa s) &  $\eta_{p}$ (Pa s)& $u_{c}$ (mm s$^{-1}$) & $\rho$ (kg m$^{-3}$) & $\lambda$  (s) & $De$ & $\tilde{\beta}$ & $\alpha$ \\
   5 &  1  & 0.6& 0.4 & 5 & 1& $0-0.3$ & $0-0.3$& 0.4 & 4\\
    \end{tabular}
    \begin{tabular}{cccccccc}
& $\epsilon$ & $h_{0}$ (mm)  & $h_{\ell}$ (mm) & $q$ (m${^2}$ s$^{-1}$) & $p_{c}$ (Pa)  & $Wi$ & $ Re$\\
Case I & 0.02& 0.4& 0.1 & $10^{-6}$ &2500 &$0-15$ & $5\times10^{-7}$ \\
Case II &0.1 & 2& 0.5 & $5\times10^{-6}$ & 100 & $0-3$ & $2.5\times10^{-6}$ \\
\end{tabular}
  \caption{Values of physical parameters used in the two-dimensional numerical simulations of the pressure-driven flow of the Oldroyd-B fluid in a hyperbolic contracting channel. The Reynolds number $Re$ is defined as $Re=\rho u_{c}h_{\ell}/\eta_{0}$ and the characteristic pressure $p_{c}$ is given as $p_{c}=\eta_{0}u_{c}\ell/h_{\ell}^{2}$.}
  \label{T2}
  \end{center}
\end{table}
In this appendix we describe the numerical techniques used to solve the system of equations (\ref{Continuity+Momentum}), (\ref{Stress tensor}), and (\ref{Evolution tau_p}). We have performed two-dimensional finite-element numerical simulations using the viscoelastic flow module in COMSOL Multiphysics, which includes the Oldroyd-B constitutive model (version 5.6, COMSOL AB, Stockholm, Sweden). All the equations are written in weak form by means of the corresponding integral scalar product, defined in terms of test functions for the pressure, velocity and polymer stress fields, i.e. $\tilde{p}$, $\tilde{u}$ and $\tilde{\tau}_{p}$, respectively. Additional details of the finite-element formulation and weak form implementation for the Oldroyd-B model in COMSOL Multiphysics are given in \citet{craven2006stabilised} and \citet{rajagopal2016analyses}.

The symmetry of the channel allows us to simplify the problem to consider only half of the domain, as shown in figure \ref{F9}. We impose the no-slip and the no-penetration boundary conditions along the wall, $y=h(z)$, and symmetry boundary condition along the centreline, $y=0$. As we are interested in determining the pressure drop $\Delta p$ originating from the contraction geometry, we have added two straight regions of length $\ell$ to eliminate the entrance and exit effects. Thus, we impose fully developed unidirectional Poiseuille velocity profile at the entrance and exit. In addition, at the inlet, we impose the polymer stress distribution corresponding to the Poiseuille flow. At the exit, the reference value for the pressure is set to zero on $y=0$. Finally, we calculate the pressure drop along the centerline between the inlet ($z=0$) and outlet ($z=\ell$) of the contraction, i.e., $\Delta p=p(y=0,z=0)-p(y=0,z=\ell)$.

We summarize in table \ref{T2}  the values of physical and geometrical parameters used in the numerical simulations. We mainly consider two hyperbolic geometries, which have an identical inlet-to-outlet ratio $\alpha=4$ but different aspect ratios $\epsilon=h_{\ell}/\ell$: $\epsilon=0.02$ (case I) and $\epsilon=0.1$ (case II). In both cases, we keep $\ell=5$ mm and $u_{c}=5$ mm $\mathrm{s}^{-1}$, while setting $h_{\ell}=0.1$ mm (case I) and $h_{\ell}=0.5$ mm (case II), and adjusting the flow rate per unit depth $q=2h_{\ell}u_{c}$, accordingly. For each case, to the study the effect of different Deborah numbers, we change the relaxation time $\lambda$ from 0 to 0.3 s to change $De$ from 0 to 0.3, while keeping all other physical and geometrical parameters. When investigating the effect of the inlet-to-outlet ratio $\alpha=h_{0}/h_{\ell}$ on the pressure drop, we change only the value of $h_{0}$ for each of the cases I and II and set $\lambda=0.2$ s, corresponding to $De=0.2$, while keeping the values of all other parameters. Similarly, when studying the effect of the polymer-to-solvent viscosity ratio $\eta_{p}/\eta_{s}$ on the pressure drop, we change only the value of $\eta_{p}$ for each of the cases I and II, while setting $\eta_{s}=1$ Pa s and $\lambda=0.1$ s ($De=0.1$), and keeping the values of all other parameters. We note that while the steady momentum equations in COMSOL Multiphysics have a convective term, the effect of fluid inertia is negligible in our simulations, as the Reynolds number is vanishingly small; see table \ref{T2}.

We discretized the velocity field using the second-order Lagrange elements and the pressure and polymer stress fields using the first-order Lagrange elements, resulting in meshes of $\approx84000$ elements for $\epsilon=0.02$ and $\approx17000$ elements for $\epsilon=0.1$. We performed tests to assess the grid sensitivity at this resolution and established grid independence. Finally, the PARDISO solver implemented in COMSOL Multiphysics has been used for simulation and the relative tolerance of the nonlinear method is always set to $10^{-5}$.

\bibliographystyle{jfm}
\bibliography{literature}

\begin{thebibliography}{59}
\expandafter\ifx\csname natexlab\endcsname\relax\def\natexlab#1{#1}\fi
\def\au#1{#1} \def\ed#1{#1} \def\yr#1{#1}\def\at#1{#1}\def\jt#1{\textit{#1}}
  \def\bt#1{#1}\def\bvol#1{\textbf{#1}} \def\vol#1{#1} \def\pg#1{#1}
  \def\publ#1{#1}\def\arxiv#1{#1}\def\org#1{#1}\def\st#1{\textit{#1}}

\bibitem[Aboubacar {\em et~al.\/}(2002)Aboubacar, Matallah \&
  Webster]{aboubacar2002highly}
{\sc \au{Aboubacar, M.}, \au{Matallah, H.} \& \au{Webster, M.~F.}} \yr{2002}
  \at{Highly elastic solutions for {O}ldroyd-{B} and {P}han-{T}hien/{T}anner
  fluids with a finite volume/element method: planar contraction flows}.
  \jt{J. Non-Newtonian Fluid Mech.}  \bvol{103}~(1),  \pg{65--103}.

\bibitem[Afonso {\em et~al.\/}(2011)Afonso, Oliveira, Pinho \&
  Alves]{afonso2011dynamics}
{\sc \au{Afonso, A.~M.}, \au{Oliveira, P.~J.}, \au{Pinho, F.~T.} \& \au{Alves,
  M.~A.}} \yr{2011}  \at{Dynamics of high-{D}eborah-number entry flows: a
  numerical study}.  \jt{J. Fluid Mech.}  \bvol{677},  \pg{272--304}.

\bibitem[Aguayo {\em et~al.\/}(2008)Aguayo, Tamaddon-Jahromi \&
  Webster]{aguayo2008excess}
{\sc \au{Aguayo, J.~P.}, \au{Tamaddon-Jahromi, H.~R.} \& \au{Webster, M.~F.}}
  \yr{2008}  \at{Excess pressure-drop estimation in contraction and expansion
  flows for constant shear-viscosity, extension strain-hardening fluids}.
  \jt{J. Non-Newtonian Fluid Mech.}  \bvol{153}~(2-3),  \pg{157--176}.

\bibitem[Ahmed \& Biancofiore(2021)]{ahmed2021new}
{\sc \au{Ahmed, H.} \& \au{Biancofiore, L.}} \yr{2021}  \at{A new approach for
  modeling viscoelastic thin film lubrication}.  \jt{J. Non-Newtonian Fluid
  Mech.}  \bvol{292},  \pg{104524}.

\bibitem[Allmendinger {\em et~al.\/}(2014)Allmendinger, Fischer, Huwyler,
  Mahler, Schwarb, Zarraga \& Mueller]{allmendinger2014rheological}
{\sc \au{Allmendinger, A.}, \au{Fischer, S.}, \au{Huwyler, J.}, \au{Mahler,
  H.~C.}, \au{Schwarb, E.}, \au{Zarraga, I.~E.} \& \au{Mueller, R.}} \yr{2014}
  \at{Rheological characterization and injection forces of concentrated protein
  formulations: {A}n alternative predictive model for non-{N}ewtonian
  solutions}.  \jt{Eur. J. Pharm. Biopharm.}  \bvol{87}~(2),  \pg{318--328}.

\bibitem[Alves {\em et~al.\/}(2003)Alves, Oliveira \&
  Pinho]{alves2003benchmark}
{\sc \au{Alves, M.~A.}, \au{Oliveira, P.~J.} \& \au{Pinho, F.~T.}} \yr{2003}
  \at{Benchmark solutions for the flow of {O}ldroyd-{B} and {PTT} fluids in
  planar contractions}.  \jt{J. Non-Newtonian Fluid Mech.}  \bvol{110}~(1),
  \pg{45--75}.

\bibitem[Alves {\em et~al.\/}(2021)Alves, Oliveira \&
  Pinho]{alves2021numerical}
{\sc \au{Alves, M.~A.}, \au{Oliveira, P.~J.} \& \au{Pinho, F.~T.}} \yr{2021}
  \at{Numerical methods for viscoelastic fluid flows}.  \jt{Annu. Rev. Fluid
  Mech.}  \bvol{53},  \pg{509--541}.

\bibitem[Binding {\em et~al.\/}(2006)Binding, Phillips \&
  Phillips]{binding2006contraction}
{\sc \au{Binding, D.~M.}, \au{Phillips, P.~M.} \& \au{Phillips, T.~N.}}
  \yr{2006}  \at{Contraction/expansion flows: The pressure drop and related
  issues}.  \jt{J. Non-Newtonian Fluid Mech.}  \bvol{137}~(1-3),  \pg{31--38}.

\bibitem[Bird {\em et~al.\/}(1987)Bird, Armstrong \&
  Hassager]{bird1987dynamics1}
{\sc \au{Bird, R.~B.}, \au{Armstrong, R.~C.} \& \au{Hassager, O.}} \yr{1987}
  {\em Dynamics of {P}olymeric {L}iquids, volume 1: {F}luid {M}echanics\/}, 2nd
  edn.  \publ{John Wiley and Sons}.

\bibitem[Boyko \& Stone(2021)]{boyko2021RT}
{\sc \au{Boyko, E.} \& \au{Stone, H.~A.}} \yr{2021}  \at{Reciprocal theorem for
  calculating the flow rate--pressure drop relation for complex fluids in
  narrow geometries}.  \jt{Phys. Rev. Fluids}  \bvol{6},  \pg{L081301}.

\bibitem[Campo-Dea{\~n}o {\em et~al.\/}(2011)Campo-Dea{\~n}o, Galindo-Rosales,
  Pinho, Alves \& Oliveira]{campo2011flow}
{\sc \au{Campo-Dea{\~n}o, L.}, \au{Galindo-Rosales, F.~J.}, \au{Pinho, F.~T.},
  \au{Alves, M.~A.} \& \au{Oliveira, M. S.~N.}} \yr{2011}  \at{Flow of low
  viscosity {B}oger fluids through a microfluidic hyperbolic contraction}.
  \jt{J. Non-Newtonian Fluid Mech.}  \bvol{166}~(21-22),  \pg{1286--1296}.

\bibitem[Chilcott \& Rallison(1988)]{chilcott1988creeping}
{\sc \au{Chilcott, M.~D.} \& \au{Rallison, J.~M.}} \yr{1988}  \at{Creeping flow
  of dilute polymer solutions past cylinders and spheres}.  \jt{J.
  Non-Newtonian Fluid Mech.}  \bvol{29},  \pg{381--432}.

\bibitem[Craven {\em et~al.\/}(2006)Craven, Rees \&
  Zimmerman]{craven2006stabilised}
{\sc \au{Craven, T.~J.}, \au{Rees, J.~M.} \& \au{Zimmerman, W.~B.}} \yr{2006}
  Stabilised finite element modelling of {O}ldroyd-{B} viscoelastic flows.
  \bt{In {\em COMSOL Conference\/}}.

\bibitem[{Datta} {\em et~al.\/}(2021){Datta}, {Ardekani}, {Arratia}, {Beris},
  {Bischofberger}, {Eggers}, {L{\'o}pez-Aguilar}, {Fielding}, {Frishman},
  {Graham}, {Guasto}, {Haward}, {Hormozi}, {McKinley}, {Poole}, {Morozov},
  {Shankar}, {Shaqfeh}, {Shen}, {Stark}, {Steinberg}, {Subramanian} \&
  {Stone}]{datta2021perspectives}
{\sc \au{{Datta}, S.~S.}, \au{{Ardekani}, A.~M.}, \au{{Arratia}, P.~E.},
  \au{{Beris}, A.~N.}, \au{{Bischofberger}, I.}, \au{{Eggers}, J.~G.},
  \au{{L{\'o}pez-Aguilar}, J.~E.}, \au{{Fielding}, S.~M.}, \au{{Frishman}, A.},
  \au{{Graham}, M.~D.}, \au{{Guasto}, J.~S.}, \au{{Haward}, S.~J.},
  \au{{Hormozi}, S.}, \au{{McKinley}, G.~H.}, \au{{Poole}, R.~J.},
  \au{{Morozov}, A.}, \au{{Shankar}, V.}, \au{{Shaqfeh}, E. S.~G.}, \au{{Shen},
  A.~Q.}, \au{{Stark}, H.}, \au{{Steinberg}, V.}, \au{{Subramanian}, G.} \&
  \au{{Stone}, H.~A.}} \yr{2021}  \at{Perspectives on viscoelastic flow
  instabilities and elastic turbulence}.  \jt{arXiv preprint arXiv:2108.09841}
  .

\bibitem[Debbaut {\em et~al.\/}(1988)Debbaut, Marchal \&
  Crochet]{debbaut1988numerical}
{\sc \au{Debbaut, B.}, \au{Marchal, J.~M.} \& \au{Crochet, M.~J.}} \yr{1988}
  \at{Numerical simulation of highly viscoelastic flows through an abrupt
  contraction}.  \jt{J. Non-Newtonian Fluid Mech.}  \bvol{29},  \pg{119--146}.

\bibitem[Fischer {\em et~al.\/}(2015)Fischer, Schmidt, Bryant \&
  Besheer]{fischer2015calculation}
{\sc \au{Fischer, I.}, \au{Schmidt, A.}, \au{Bryant, A.} \& \au{Besheer, A.}}
  \yr{2015}  \at{Calculation of injection forces for highly concentrated
  protein solutions}.  \jt{Int. J. Pharm.}  \bvol{493}~(1-2),  \pg{70--74}.

\bibitem[Groisman \& Quake(2004)]{groisman2004microfluidic}
{\sc \au{Groisman, A.} \& \au{Quake, S.~R.}} \yr{2004}  \at{A microfluidic
  rectifier: anisotropic flow resistance at low {R}eynolds numbers}.  \jt{Phys.
  Rev. Lett.}  \bvol{92}~(9),  \pg{094501}.

\bibitem[Groisman \& Steinberg(1996)]{groisman1996couette}
{\sc \au{Groisman, A.} \& \au{Steinberg, V.}} \yr{1996}  \at{Couette-{T}aylor
  flow in a dilute polymer solution}.  \jt{Phys. Rev. Lett.}  \bvol{77}~(8),
  \pg{1480}.

\bibitem[Hsiao {\em et~al.\/}(2017)Hsiao, Dinic, Ren, Sharma \&
  Schroeder]{hsiao2017passive}
{\sc \au{Hsiao, K.~W.}, \au{Dinic, J.}, \au{Ren, Y.}, \au{Sharma, V.} \&
  \au{Schroeder, C.~M.}} \yr{2017}  \at{Passive non-linear microrheology for
  determining extensional viscosity}.  \jt{Phys. Fluids}  \bvol{29}~(12),
  \pg{121603}.

\bibitem[James(2016)]{james2016n1}
{\sc \au{James, D.~F.}} \yr{2016}  \at{N1 stresses in extensional flows}.
  \jt{J. Non-Newtonian Fluid Mech.}  \bvol{232},  \pg{33--42}.

\bibitem[Keiller(1993)]{keiller1993entry}
{\sc \au{Keiller, R.~A.}} \yr{1993}  \at{Entry-flow calculations for the
  {O}ldroyd-{B} and {FENE} equations}.  \jt{J. Non-Newtonian Fluid Mech.}
  \bvol{46}~(2-3),  \pg{143--178}.

\bibitem[Keshavarz \& McKinley(2016)]{keshavarz2016micro}
{\sc \au{Keshavarz, B.} \& \au{McKinley, G.~H.}} \yr{2016}  \at{Micro-scale
  extensional rheometry using hyperbolic converging/diverging channels and jet
  breakup}.  \jt{Biomicrofluidics}  \bvol{10}~(4),  \pg{043502}.

\bibitem[Keunings(2004)]{keunings2004micro}
{\sc \au{Keunings, R.}} \yr{2004}  \at{Micro-macro methods for the multiscale
  simulation of viscoelastic flow using molecular models of kinetic theory}.
  \jt{Rheol. Rev.}  \bvol{2004},  \pg{67--98}.

\bibitem[Koppol {\em et~al.\/}(2009)Koppol, Sureshkumar, Abedijaberi \&
  Khomami]{koppol2009anomalous}
{\sc \au{Koppol, A.~P.}, \au{Sureshkumar, R.}, \au{Abedijaberi, A.} \&
  \au{Khomami, B.}} \yr{2009}  \at{Anomalous pressure drop behaviour of mixed
  kinematics flows of viscoelastic polymer solutions: a multiscale simulation
  approach}.  \jt{J. Fluid Mech.}  \bvol{631},  \pg{231}.

\bibitem[Larson(1992)]{larson1992instabilities}
{\sc \au{Larson, R.~G.}} \yr{1992}  \at{Instabilities in viscoelastic flows}.
  \jt{Rheol. Acta}  \bvol{31}~(3),  \pg{213--263}.

\bibitem[L{\'o}pez-Aguilar {\em et~al.\/}(2016)L{\'o}pez-Aguilar,
  Tamaddon-Jahromi, Webster \& Walters]{lopez2016numerical}
{\sc \au{L{\'o}pez-Aguilar, J.~E.}, \au{Tamaddon-Jahromi, H.~R.}, \au{Webster,
  M.~F.} \& \au{Walters, K.}} \yr{2016}  \at{Numerical vs experimental pressure
  drops for {B}oger fluids in sharp-corner contraction flow}.  \jt{Phys.
  Fluids}  \bvol{28}~(10),  \pg{103104}.

\bibitem[Morozov \& Spagnolie(2015)]{Intro_C_F}
{\sc \au{Morozov, A.} \& \au{Spagnolie, S.~E.}} \yr{2015}  \at{Introduction to
  complex fluids}.  \bt{In {\em Complex Fluids in Biological Systems\/} (ed.
  \ed{S.~E. Spagnolie})},  \pg{pp. 3--52}.  \publ{Springer}.

\bibitem[Nguyen {\em et~al.\/}(2008)Nguyen, Lam, Ho \&
  Low]{nguyen2008improvement}
{\sc \au{Nguyen, N.~T.}, \au{Lam, Y.~C.}, \au{Ho, S.S.} \& \au{Low, C. L.~N.}}
  \yr{2008}  \at{Improvement of rectification effects in diffuser/nozzle
  structures with viscoelastic fluids}.  \jt{Biomicrofluidics}  \bvol{2}~(3),
  \pg{034101}.

\bibitem[Nigen \& Walters(2002)]{nigen2002viscoelastic}
{\sc \au{Nigen, S.} \& \au{Walters, K.}} \yr{2002}  \at{Viscoelastic
  contraction flows: comparison of axisymmetric and planar configurations}.
  \jt{J. Non-Newtonian Fluid Mech.}  \bvol{102}~(2),  \pg{343--359}.

\bibitem[Nystr{\"o}m {\em et~al.\/}(2012)Nystr{\"o}m, Tamaddon-Jahromi, Stading
  \& Webster]{nystrom2012numerical}
{\sc \au{Nystr{\"o}m, M.}, \au{Tamaddon-Jahromi, H.~R.}, \au{Stading, M.} \&
  \au{Webster, M.~F.}} \yr{2012}  \at{Numerical simulations of {B}oger fluids
  through different contraction configurations for the development of a
  measuring system for extensional viscosity}.  \jt{Rheol. Acta}
  \bvol{51}~(8),  \pg{713--727}.

\bibitem[Nystr{\"o}m {\em et~al.\/}(2016)Nystr{\"o}m, Tamaddon-Jahromi, Stading
  \& Webster]{nystrom2016extracting}
{\sc \au{Nystr{\"o}m, M.}, \au{Tamaddon-Jahromi, H.~R.}, \au{Stading, M.} \&
  \au{Webster, M.~F.}} \yr{2016}  \at{Extracting extensional properties through
  excess pressure drop estimation in axisymmetric contraction and expansion
  flows for constant shear viscosity, extension strain-hardening fluids}.
  \jt{Rheol. Acta}  \bvol{55}~(5),  \pg{373--396}.

\bibitem[Nystr{\"o}m {\em et~al.\/}(2017)Nystr{\"o}m, Tamaddon-Jahromi, Stading
  \& Webster]{nystrom2017hyperbolic}
{\sc \au{Nystr{\"o}m, M.}, \au{Tamaddon-Jahromi, H.~R.}, \au{Stading, M.} \&
  \au{Webster, M.~F.}} \yr{2017}  \at{Hyperbolic contraction measuring systems
  for extensional flow}.  \jt{Mech. Time-Dependent Mater.}  \bvol{21}~(3),
  \pg{455--479}.

\bibitem[Ober {\em et~al.\/}(2013)Ober, Haward, Pipe, Soulages \&
  McKinley]{ober2013microfluidic}
{\sc \au{Ober, T.~J.}, \au{Haward, S.~J.}, \au{Pipe, C.~J.}, \au{Soulages, J.}
  \& \au{McKinley, G.~H.}} \yr{2013}  \at{Microfluidic extensional rheometry
  using a hyperbolic contraction geometry}.  \jt{Rheol. Acta}  \bvol{52}~(6),
  \pg{529--546}.

\bibitem[Oliveira {\em et~al.\/}(2007)Oliveira, Oliveira, Pinho \&
  Alves]{oliveira2007effect}
{\sc \au{Oliveira, M. S.~N.}, \au{Oliveira, P.~J.}, \au{Pinho, F.~T.} \&
  \au{Alves, M.~A.}} \yr{2007}  \at{Effect of contraction ratio upon
  viscoelastic flow in contractions: the axisymmetric case}.  \jt{J.
  Non-Newtonian Fluid Mech.}  \bvol{147}~(1-2),  \pg{92--108}.

\bibitem[Owens \& Phillips(2002)]{owens2002computational}
{\sc \au{Owens, R.~G.} \& \au{Phillips, T.~N.}} \yr{2002} {\em Computational
  rheology\/}.  \publ{Imperial College Press}.

\bibitem[Pearson(1985)]{pearson}
{\sc \au{Pearson, J. R.~A.}} \yr{1985} {\em Mechanics of {P}olymer
  {P}rocessing\/}.  \publ{Elsevier}.

\bibitem[P{\'e}rez-Salas {\em et~al.\/}(2019)P{\'e}rez-Salas, S{\'a}nchez,
  Ascanio \& Aguayo]{perez2019analytical}
{\sc \au{P{\'e}rez-Salas, K.~Y.}, \au{S{\'a}nchez, S.}, \au{Ascanio, G.} \&
  \au{Aguayo, J.~P.}} \yr{2019}  \at{Analytical approximation to the flow of a
  sptt fluid through a planar hyperbolic contraction}.  \jt{J. Non-Newtonian
  Fluid Mech.}  \bvol{272},  \pg{104160}.

\bibitem[Phan-Thien(1978)]{phan1978nonlinear}
{\sc \au{Phan-Thien, N.}} \yr{1978}  \at{A nonlinear network viscoelastic
  model}.  \jt{J. Rheol.}  \bvol{22}~(3),  \pg{259--283}.

\bibitem[Phan-Thien \& Tanner(1977)]{thien1977new}
{\sc \au{Phan-Thien, N.} \& \au{Tanner, R.~I.}} \yr{1977}  \at{A new
  constitutive equation derived from network theory}.  \jt{J. Non-Newtonian
  Fluid Mech.}  \bvol{2}~(4),  \pg{353--365}.

\bibitem[Rajagopal \& Das(2016)]{rajagopal2016analyses}
{\sc \au{Rajagopal, M.~C.} \& \au{Das, S.~K.}} \yr{2016}  \at{Analyses of drag
  on viscoelastic liquid infused bio-inspired patterned surfaces}.  \jt{J.
  Non-Newtonian Fluid Mech.}  \bvol{228},  \pg{17--30}.

\bibitem[Rothstein \& McKinley(1999)]{rothstein1999extensional}
{\sc \au{Rothstein, J.~P.} \& \au{McKinley, G.~H.}} \yr{1999}  \at{Extensional
  flow of a polystyrene {B}oger fluid through a 4: 1: 4 axisymmetric
  contraction/expansion}.  \jt{J. Non-Newtonian Fluid Mech.}  \bvol{86}~(1-2),
  \pg{61--88}.

\bibitem[Rothstein \& McKinley(2001)]{rothstein2001axisymmetric}
{\sc \au{Rothstein, J.~P.} \& \au{McKinley, G.~H.}} \yr{2001}  \at{The
  axisymmetric contraction--expansion: the role of extensional rheology on
  vortex growth dynamics and the enhanced pressure drop}.  \jt{J. Non-Newtonian
  Fluid Mech.}  \bvol{98}~(1),  \pg{33--63}.

\bibitem[Saprykin {\em et~al.\/}(2007)Saprykin, Koopmans \&
  Kalliadasis]{saprykin2007free}
{\sc \au{Saprykin, S.}, \au{Koopmans, R.~J.} \& \au{Kalliadasis, S.}} \yr{2007}
   \at{Free-surface thin-film flows over topography: influence of inertia and
  viscoelasticity}.  \jt{J. Fluid Mech.}  \bvol{578},  \pg{271--293}.

\bibitem[Shaqfeh(1996)]{shaqfeh1996purely}
{\sc \au{Shaqfeh, E. S.~G.}} \yr{1996}  \at{Purely elastic instabilities in
  viscometric flows}.  \jt{Annu. Rev. Fluid Mech.}  \bvol{28}~(1),
  \pg{129--185}.

\bibitem[Sousa {\em et~al.\/}(2009)Sousa, Coelho, Oliveira \&
  Alves]{sousa2009three}
{\sc \au{Sousa, P.~C.}, \au{Coelho, P.~M.}, \au{Oliveira, M. S.~N.} \&
  \au{Alves, M.~A.}} \yr{2009}  \at{Three-dimensional flow of {N}ewtonian and
  {B}oger fluids in square--square contractions}.  \jt{J. Non-Newtonian Fluid
  Mech.}  \bvol{160}~(2-3),  \pg{122--139}.

\bibitem[Sousa {\em et~al.\/}(2010)Sousa, Pinho, Oliveira \&
  Alves]{sousa2010efficient}
{\sc \au{Sousa, P.~C.}, \au{Pinho, F.~T.}, \au{Oliveira, M. S.~N.} \&
  \au{Alves, M.~A.}} \yr{2010}  \at{Efficient microfluidic rectifiers for
  viscoelastic fluid flow}.  \jt{J. Non-Newtonian Fluid Mech.}
  \bvol{165}~(11-12),  \pg{652--671}.

\bibitem[Steinberg(2021)]{steinberg2021elastic}
{\sc \au{Steinberg, V.}} \yr{2021}  \at{Elastic {T}urbulence: {A}n
  {E}xperimental {V}iew on {I}nertialess {R}andom {F}low}.  \jt{Annu. Rev.
  Fluid Mech.}  \bvol{53},  \pg{27--58}.

\bibitem[Szabo {\em et~al.\/}(1997)Szabo, Rallison \& Hinch]{szabo1997start}
{\sc \au{Szabo, P.}, \au{Rallison, J.~M.} \& \au{Hinch, E.~J.}} \yr{1997}
  \at{Start-up of flow of a {FENE}-fluid through a 4:1:4 constriction in a
  tube}.  \jt{J. Non-Newtonian Fluid Mech.}  \bvol{72}~(1),  \pg{73--86}.

\bibitem[Tadmor \& Gogos(2013)]{tadmor2013principles}
{\sc \au{Tadmor, Z.} \& \au{Gogos, C.~G.}} \yr{2013} {\em Principles of polymer
  processing\/}.  \publ{John Wiley and Sons}.

\bibitem[Tamaddon-Jahromi {\em et~al.\/}(2016)Tamaddon-Jahromi, Gardu{\~n}o,
  L{\'o}pez-Aguilar \& Webster]{tamaddon2016predicting}
{\sc \au{Tamaddon-Jahromi, H.~R.}, \au{Gardu{\~n}o, I.~E.},
  \au{L{\'o}pez-Aguilar, J.~E.} \& \au{Webster, M.~F.}} \yr{2016}
  \at{Predicting large experimental excess pressure drops for {B}oger fluids in
  contraction--expansion flow}.  \jt{J. Non-Newtonian Fluid Mech.}  \bvol{230},
   \pg{43--67}.

\bibitem[Tamaddon-Jahromi {\em et~al.\/}(2018)Tamaddon-Jahromi,
  L{\'o}pez-Aguilar \& Webster]{tamaddon2018modelling}
{\sc \au{Tamaddon-Jahromi, H.~R.}, \au{L{\'o}pez-Aguilar, J.~E.} \&
  \au{Webster, M.~F.}} \yr{2018}  \at{On modelling viscoelastic flow through
  abrupt circular 8:1 contractions--matching experimental pressure-drops and
  vortex structures}.  \jt{J. Non-Newtonian Fluid Mech.}  \bvol{251},
  \pg{28--42}.

\bibitem[Tamaddon-Jahromi {\em et~al.\/}(2010)Tamaddon-Jahromi, Webster \&
  Walters]{tamaddon2010predicting}
{\sc \au{Tamaddon-Jahromi, H.~R.}, \au{Webster, M.~F.} \& \au{Walters, K.}}
  \yr{2010}  \at{Predicting numerically the large increases in extra pressure
  drop when {B}oger fluids flow through axisymmetric contractions}.  \jt{J.
  Nat. Sci.}  \bvol{2}~(1),  \pg{1--11}.

\bibitem[Tamaddon-Jahromi {\em et~al.\/}(2011)Tamaddon-Jahromi, Webster \&
  Williams]{jahromi2011excess}
{\sc \au{Tamaddon-Jahromi, H.~R.}, \au{Webster, M.~F.} \& \au{Williams, P.~R.}}
  \yr{2011}  \at{Excess pressure drop and drag calculations for
  strain-hardening fluids with mild shear-thinning: contraction and falling
  sphere problems}.  \jt{J. Non-Newtonian Fluid Mech.}  \bvol{166}~(16),
  \pg{939--950}.

\bibitem[Tavakol {\em et~al.\/}(2017)Tavakol, Holmes \&
  Stone]{tavakol2017extended}
{\sc \au{Tavakol, B.and~Froehlicher, G.}, \au{Holmes, D.~P.} \& \au{Stone,
  H.~A.}} \yr{2017}  \at{Extended lubrication theory: improved estimates of
  flow in channels with variable geometry}.  \jt{Proc. R. Soc. A}
  \bvol{473}~(2206),  \pg{20170234}.

\bibitem[Tichy(1996)]{tichy1996non}
{\sc \au{Tichy, J.~A.}} \yr{1996}  \at{Non-{N}ewtonian lubrication with the
  convected {M}axwell model}.  \jt{Trans. ASME J. Tribol.}  \bvol{118},
  \pg{344--348}.

\bibitem[Webster {\em et~al.\/}(2019)Webster, Tamaddon-Jahromi,
  L{\'o}pez-Aguilar \& Binding]{webster2019enhanced}
{\sc \au{Webster, M.~F.}, \au{Tamaddon-Jahromi, H.~R.}, \au{L{\'o}pez-Aguilar,
  J.~E.} \& \au{Binding, D.~M.}} \yr{2019}  \at{Enhanced pressure drop, planar
  contraction flows and continuous spectrum models}.  \jt{J. Non-Newtonian
  Fluid Mech.}  \bvol{273},  \pg{104184}.

\bibitem[White \& Metzner(1963)]{white1963development}
{\sc \au{White, J.~L.} \& \au{Metzner, A.~B.}} \yr{1963}  \at{Development of
  constitutive equations for polymeric melts and solutions}.  \jt{J. Appl.
  Polym. Sci.}  \bvol{7}~(5),  \pg{1867--1889}.

\bibitem[Zhang {\em et~al.\/}(2002)Zhang, Matar \&
  Craster]{zhang2002surfactant}
{\sc \au{Zhang, Y.~L.}, \au{Matar, O.~K.} \& \au{Craster, R.~V.}} \yr{2002}
  \at{Surfactant spreading on a thin weakly viscoelastic film}.  \jt{J.
  Non-Newtonian Fluid Mech.}  \bvol{105}~(1),  \pg{53--78}.

\bibitem[Zografos {\em et~al.\/}(2020)Zografos, Hartt, Hamersky, Oliveira,
  Alves \& Poole]{zografos2020viscoelastic}
{\sc \au{Zografos, K.}, \au{Hartt, W.}, \au{Hamersky, M.}, \au{Oliveira, M.
  S.~N.}, \au{Alves, M.~A.} \& \au{Poole, R.~J.}} \yr{2020}  \at{Viscoelastic
  fluid flow simulations in the e-{VROC}\textsuperscript{TM} geometry}.  \jt{J.
  Non-Newtonian Fluid Mech.}  \bvol{278},  \pg{104222}.

\end{thebibliography}


\end{document}